\documentclass[12pt]{article}
\usepackage{amsmath}
\PassOptionsToPackage{righttag}{amsmath}
\def\myline{\hbox to \textwidth}
\def\eqn#1{\hbox{(\ref{#1})}}

\def\mtrx#1{\begin{matrix}#1\end{matrix}}
\def\pmtrx#1{\begin{pmatrix}#1\end{pmatrix}}
\def\bra{\langle}
\def\ket{\rangle}
\def\ftm{\phantom}
\def\undl{\underline}
\def\mlt{\multicolumn}
\def\dst{\displaystyle}
\def\tst{\textstyle}

\def\cnj{\,\tst\ast}
\def\vct#1{\mbox{\boldmath $#1$}}
\def\buildchar#1#2#3{{\null\!                   
   \mathop{\vphantom{#1}\smash#1}\limits
   ^{#2}_{#3}
   \!\null}}                                    

\def\onedot#1{\buildchar{#1}{\dst.}{}}
\def\twodots#1{\buildchar{#1}{\dst..}{}}

\catcode`@=11
\def\eqalign#1{\null\,\vcenter{\openup\jot\m@th
  \ialign{\strut\hfil$\displaystyle{##}$&$\displaystyle{{}##}$\hfil
      \crcr#1\crcr}}\,}
\def\eqalignno#1{\displ@y \tabskip\@centering
  \halign to\displaywidth{\hfil$\@lign\displaystyle{##}$\tabskip\z@skip
    &$\@lign\displaystyle{{}##}$\hfil\tabskip\@centering
    &\llap{$\@lign##$}\tabskip\z@skip\crcr
    #1\crcr}}
\def\leqalignno#1{\displ@y \tabskip\@centering
  \halign to\displaywidth{\hfil$\@lign\displaystyle{##}$\tabskip\z@skip
    &$\@lign\displaystyle{{}##}$\hfil\tabskip\@centering
    &\kern-\displaywidth\rlap{$\@lign##$}\tabskip\displaywidth\crcr
    #1\crcr}}

\long\def\@makefntext#1{
  \vskip0pt\parindent0pt\begin{list}{}%
  {\labelwidth1.5em\leftmargin=\labelwidth%
     \labelsep3pt\itemsep0pt\parsep0pt\topsep-2pt
            \def\baselinestretch{1.0}\footnotesize}%
  \item[\hfill\@makefnmark]#1\end{list}}

\long\def\@makecaption#1#2{\vskip10pt
    {#1:\ \begingroup\small\baselineskip14pt plus4pt minus2pt
             #2\par\endgroup}}

\catcode`@=12
\def\bln#1#2\eln{\begin{equation}\label{#1}
            \eqalign{#2}\end{equation}\noindent}
\def\blb#1#2\elb{\begin{equation}\label{#1}%
            \left\{\eqalign{#2}\right.\end{equation}\noindent}
\def\bqx#1#2\eqx{$$\eqalignno{\refstepcounter{equation}\label{#1}#2}$$}

\def\mylbl#1{(\theequation\hbox{$#1$})}

\def\eql#1#2{\hbox{(\ref{#1}$#2$)}}
\def\blgn#1#2#3\elgn{\def\minalignsep{#1}
   \begin{equation}\label{#2}\begin{aligned}#3\end{aligned}\end{equation}}
\def\blgb#1#2#3\elgb{\def\minalignsep{#1}
   \begin{equation}\label{#2}\left\{\begin{aligned}#3\end{aligned}\right.\end{equation}}
\def\jtem#1#2{\par\hangafter0\hangindent#1
              \noindent\llap{#2\enspace}\ignorespaces}

\def\thesection{\arabic{section}}
\def\theequation{\thesection.\arabic{equation}}
\def\sectn#1#2{\section{#1\label{#2}}\setcounter{equation}{0}}
\def\appendx{\setcounter{section}{0}
	\def\thesection{\Alph{section}}
		\def\theequation{\Alph{section}.\arabic{equation}}}
\def\vsp{\hbox{\vrule height12.5pt depth3.5pt width0pt}}
\def\indnt{\vskip0pt\noindent\vsp\hskip1.5em}
\def\brlist{\begin{thebibliography}{999}}
\def\brf{\bibitem}
\def\erlist{\end{thebibliography}}
\begin{document}
\thispagestyle{empty}
\begin{center}{\sf TU PREPRINT}\end{center}
\vglue 2mm
\rightline{\vtop{\hbox{TU-E18-06-0801}\hbox{tupaper11.tex}%
}}%
\vskip2cm%
\begin{center}{\Large\bf
Covariant Helicity-Coupling Amplitudes:\\
A New Formulation\\[3mm]
}
\vskip24pt%
Suh-Urk Chung$^{\ *}$ and Jan Michael Friedrich
\vskip 2pt%
{\em Physik Department E18, Technische Universit\"at M\"unchen, Germany}
\vskip1cm%
\today
\vskip2cm%
{\large\bf abstract}
\vskip1cm%
\begin{quote}
\indnt
We have worked out covariant amplitudes for any two-body decay
of a resonance with an arbitrary non-zero mass, 
which involves arbitrary integer spins in the initial and the final states. 
One key new ingredient  for this work is the application of the total intrinsic spin
operator $\vec S$ which is given directly in terms of the generators of the Poincar\'e
group.
\indnt
Using the results of this study, we show how to explore the Lorentz factors 
which appear naturally, if the momentum-space wave functions are used 
to form the covariant decay amplitudes.  We have devised a method of constructing
our covariant decay amplitudes, such that they lead to the Zemach amplitudes when
the Lorentz factors are set one.
\end{quote}
\end{center}
\vspace*{\fill}\footnoterule\vskip-10mm\footins{\small
\hskip3pt\jtem{2pt}{$^*\!\!$}{Mercator Visiting Professor for 2003--2005 and 2007--2009\\
Permanent Address: \vtop{\hbox{Physics Department, Brookhaven National Laboratory,}
		\hbox{Upton, NY 11973, USA}}}}
\eject
\pagenumbering{arabic}
\def\baselinestretch{1.4}\normalsize
\pagenumbering{arabic}
%
\sectn{Introduction}{sc1}
\indnt
   When one attempts to write down the covariant amplitudes for two-body decay
processes involving arbitrary spin, one discovers invariably that the Lorentz factors $\gamma$
(defined as the energy for each of the decay products 
divided by its invariant mass) appear in the amplitudes.  
This is an endemic feature of any phenomenological amplitude one writes down in four-momentum
space. The purpose of this paper is to systematically investigate functional
dependence of such Lorentz factors.

   In previous publications and notes by one of the authors\cite{SCh0}\cite{SCh1},
a few of the more relevant examples of this problem have been worked out, but it was all done
under the assumption that the total intrinsic spin can be derived from `fictitious'
wave functions of the decay products, which are deemed to be {\em at rest} in the parent
rest frame.  We show in this paper how this problem could be properly handled in the general
relativistic frame work---the key ideas involved are succinctly summarized in Section 2.
Consider the canonical state $|\vec p,jm\ket$ of a single particle.  A rotation operator
which acts on the state modifies {\em both} $\vec p$ and $|jm\ket$, the prescription of which
is very well known and it involves an introduction of angular momentum operators $\vec J$
in the expression for the rotation operators.  
There exists, however, a {\em new subgroup of rotations}
of $SU(2)$ which leaves $\vec p$ invariant, affecting only the $|jm\ket$.  This is accomplished
by replacing $\vec J$ in the rotation operator with $\vec S$, but now the $\vec S$ becomes
a complicated function of {\em all the generators} of the Poincar\'e group, 
i.e. $\vec J$ (angular momentum operator),
$\vec K$ (the boost operator) and $\vec P$ (the translation operator).  It has been shown that
the rotation operator involving $\vec L=\vec J-\vec S$ instead of $\vec S$ affects $\vec p$
while leaving $|jm\ket$ invariant\cite{McK}\cite{Mcf}.  The operators $\vec S$ and $\vec L$
provide fully relativistic prescription for the familiar $LS$ coupling scheme for two-particle
final states.  The application of this concept to the problem of exploring the functional
dependence of the Lorentz factors leads to the generators of the Lorentz group in four-momentum
space, in which the operators $\vec P$ is replaced by their eigenvalues $\vec p$.

   In Sections 3 and 4, 
we start from general wave functions at rest of arbitrary integer spin
and a description of the kinematics of the two-body decay; much of these results have been
worked out by one of us in a previous publication\cite{SCh1}, but they are reproduced
here for ease of reference.  We cover in Section 5 the general wave functions in a given
total intrinsic spin.  In Section 6, we  cover the rules 
for constructing covariant helicity-coupling amplitudes---this constitutes the heart of
what is new in this paper.  A number of illustrative examples are given in Sections 7, 8 and 9,
for the decay amplitudes in which decay products have spin 0, 1 or 2.
We have worked out in Section 10 an example of a sequence of decay chains: a parent
state of spin $J$ decays into two intermediate states of spins $s$ and $\sigma$, each of which
decays into $2\pi$ and $3\pi$, respectively.  
We have given in some detail a possible set of angles defined in two helicity coordinate systems,
the rest frames of $s$ and $\sigma$, which are needed to describe the secondary decays.
Section 11 is reserved for conclusions and discussions.  

The Appendix A deals with
the generators of the Lorentz group, starting with the general rank-2 angular momentum
tensor in four-momentum space.  In the Appendix B, we have worked out a derivation of
the spherical harmonics from the tensor wave functions $\phi(J\,m)$ given in Section 4.
The Appendix C deals with certain additional decay amplitudes
not included in the main text, to show why and how they have been excluded in the construction
of covariant decay amplitudes in our paper; 
  the excluded decay amplitudes violate the $r^\ell$ rule,
  i.e. an orbital angular momentum $\ell$ must induce an $r^\ell$ dependence
  in the amplitude, where $r$
  is the relative momentum of the daughter states in the parent rest frame.
\sectn{Relativistic Two-Body Systems}{sc2}
\indnt
   Consider a two-body decay $J\to s+\sigma$.  The {\em canonical states} for the
decay products are denoted $|\vec q,sm_s\ket$ and $|\vec k, \sigma m_\sigma\ket$, where
$\vec q$ and $\vec k$ stand for the momenta for the daughter particles.
So $\vec q=-\vec k$ in the $J$ rest frame ($J\,$RF).

   In this paper we use the word `relativistic' to mean 
that the magnitude of the momentum $q$ or $k$ can take on arbitrary values 
and, therefore, that the Lorentz factors $\gamma_s>1$ or $\gamma_\sigma>1$ 
can be arbitrarily large.  
Conversely, the word `non-relativistic' means that a statement is true only 
in the limit $q\approx0$ or $k\approx0$ and $\gamma_s\approx1$ or $\gamma_\sigma\approx1$.

   The central idea used in this paper can be summarized as follows.
Consider a relativistic canonical state for $s$, i.e. $|\vec q,sm_s\ket$.
We know how it transforms under an arbitrary rotation $R(\alpha,\beta,\gamma)$
\bln{tb1}
  R(\alpha,\beta,\gamma)\, |\,q_{\,i},\;s\,m_s\ket
	=\sum_{m'_s\,k}\,|\,R_{\,ik}\,q_{\,k},\;s\,m'_s\ket\,
	D^{\,s}_{m'_s\,m^{\ftm{\prime}}_s}(\alpha,\beta,\gamma),\quad i,k=1,\,2,\,3
\eln
where $R$ on the left-hand side is the usual rotation operator given by
\bln{tb2}
   R(\alpha,\beta,\gamma)=\exp[-i\,\alpha\,J_z]\,\exp[-i\,\beta\,J_y]\,\exp[-i\,\gamma\,J_z]
\eln
and $R_{\,ik}$ on the right-hand side is the corresponding $3\times3$ matrix.
It has been shown\cite{McK}\cite{Mcf} that the angular momentum operator $\vec J$
can be broken up into two components
\bln{tb3}
	\vec J=\vec S+\vec L
\eln
where, with {$w$} denoting the mass of a particle,
\bln{tb3a}
 w\,{\vec S}&=P^0{\vec J}-{\vec P}\times{\vec K}
	-\frac{1}{P^0+w}{\vec P}\,({\vec P}\cdot{\vec J})\cr
\eln
{$\vec J$}, {$\vec K$} and {$(P^0,\vec P)$} are 
the generators of the {Poincar\'e group}.  Here {$\vec L$} is a derived quantity,
defined by {$\vec L=\vec J-\vec S$}.  
The operators $\vec S$ and $\vec L$ satisfy
the standard Lie algebra of the rotation group and they commute
\bqx{tb3b}
   [S_i,\,S_j]&=i\,\epsilon_{ijk}\,S_k,\quad
	[L_i,\,L_j]=i\,\epsilon_{ijk} \,L_k,\quad [S_i,L_j]=0,\quad i,j,k=1,2,3&\mylbl{a}\cr
 [P^0,L_i]&=0,\quad [P_i,\,L_j]=i\,\epsilon_{ijk}\,P_k,\quad \vec P\cdot\vec L=0&\mylbl{b}
\eqx
It was shown\cite{McK} that
\bqx{tb4}
   R^S(\alpha,\beta,\gamma)\,|\,q_{\,i},\;s\,{m_s}\ket
	&=\sum_{m'_s}\,|\,q_{\,i},\;s\,m'_s\ket\,
	D^{\,s}_{m'_s\,m^{\ftm{\prime}}_s}(\alpha,\beta,\gamma)&\mylbl{a}\cr
   R^L(\alpha,\beta,\gamma)\, |\,q_{\,i},\;s\,m_s\ket
	&=\sum_{k}\,|\,R_{\,ik}\,q_{\,k},\;s\,m_s\ket&\mylbl{b}\cr
\eqx
where
\bln{tb5}
   R^S(\alpha,\beta,\gamma)
	&=\exp[-i\,\alpha\,S_z]\,\exp[-i\,\beta\,S_y]\,\exp[-i\,\gamma\,S_z]\cr
   R^L(\alpha,\beta,\gamma)
	&=\exp[-i\,\alpha\,L_z]\,\exp[-i\,\beta\,L_y]\,\exp[-i\,\gamma\,L_z]\cr
\eln
The equations \eql{tb4}{a} and \eql{tb4}{b} show that the operators $\vec S$ and $\vec L$
induce {\em separate rotations} on $\vec q$ and $|sm_s\ket$, 
which are normally transformed simultaneously in a conventional rotation [\,see \eqn{tb1}\,].
Thus $R^S(\alpha,\beta,\gamma)$ acts on the relativistic state $|\,q_{\,i},\;s\,m_s\ket$
as if it were a {\em rest state}, affecting only the spin components, whereas
$R^L(\alpha,\beta,\gamma)$ acts on the relativistic state $|\,q_{\,i},\;s\,m_s\ket$
as if it were a spinless state, leaving the spin components invariant.

   We are now ready to apply $R^S(\alpha,\beta,\gamma)$ and $R^L(\alpha,\beta,\gamma)$
to relativistic two-body systems.  Consider a system consisting of two particles
with momenta $\vec q$ and $\vec k$ with their spin states given by
$|s\,m_s\ket$  and $|\sigma\, m_\sigma\ket$.  Define
\bln{tb6}
  \vec S=\vec S_s+\vec S_\sigma\quad{\rm and}\quad\vec L=\vec L_s+\vec L_\sigma
\eln
We define a two-body system in its rest frame 
(i.e. $\vec q+\vec k=0$ and $\vec r=\vec q-\vec k\;$) by
\bln{tb7}
   |\,\vec r\;S\,m\,\ket=\sum_{m_s\,m_\sigma}\,(sm_s\;\sigma m_\sigma\,|\,S\,m\,)
	|\,\vec q,\;s\,m_s\ket\;|\,\vec k,\;\sigma\,m_\sigma\ket
\eln
We use the notation 
$(s_1m_1\,s_2m_2\,|\,S\,m_s)$ to stand for the standard Clebsch-Gordan coefficients\cite{Rose}.
  Under a rotation $R^S$, the above state transforms as
  \bqx{tb8}
  R^S(\phi,\theta,0)\,|\,r_i,\;S\,m\,\ket
  &=\sum_{m'}\,|\,r_i,\;S\,m'\,\ket\,
  D^{\,S}_{m'\,m\,}(\phi,\theta,0),\quad i=1,\,2,\,3&\mylbl{a}\cr
  R^L(\phi,\theta,0)\,|\,r_i,\;S\,m\,\ket
  &=|\,R_{ik}\,r_k,\;S\,m\,\ket,\quad i,k=1,\,2,\,3&\mylbl{b}\cr
  \eqx
  where $\Omega=(\theta,\phi)$ describes the direction of $\vec r$
  (as well as $\vec q$ and $\vec k$\,) in the $J\,$RF.
  This shows the state $|\,\vec r,\;S\,m\,\ket$ acts
  like a {\em rest state} under $R^S(\phi,\theta,0)$ in the sense that 
  $\vec r$ remains invariant.
  So we have established the concept of a total intrinsic spin $S$
  for a relativistic two-body system.  The operator $R^L(\phi,\theta,0)$
  acts only on the $\vec r$ as if the two final-state particles involved
  were spinless.
  
  It is clear that this leads naturally to the concept of orbital angular
  momentum for the relativistic two-body system 
  \bln{tb8a}
  |\,\ell\,m_\ell\;S\,m\ket\propto\int{\rm d}\Omega\,Y^\ell_{m_{\ell}}(\Omega)\,|\,\vec r,\;S\,m\,\ket
  =\sqrt{\frac{2\ell+1}{4\pi}}\int{\rm d}\Omega\,D^{\,\ell\,\cnj}_{m\,0}(\phi,\theta,0)\,|\,\vec r,\;S\,m\,\ket
  \eln 
  The analogue of \eqn{tb8} for this ket state is
  \bqx{tb8b}
  R^S(\phi,\theta,0)\,|\,\ell\,m_\ell\;S\,m\ket
  &=\sum_{m'}\,|\,\ell\,m_\ell\;S\,m'\ket\,D^S_{m'\,m}(\phi,\theta,0)&\mylbl{a}\cr
  R^L(\phi,\theta,0)\,|\,\ell\,m_\ell\;S\,m\ket
  &=\sum_{m'_\ell}\,|\,\ell\,m'_\ell\;S\,m\ket\,D^\ell_{m'_\ell\,m_\ell}(\phi,\theta,0)&\mylbl{b}
  \eqx
  The formula \eql{tb8b}{b} can be shown by substituting \eqn{tb7} 
  into \eqn{tb8a} and carrying out the actions
  of $R^L$ on the vectors $\vec q=\vec r/2$ and  $\vec k=-\vec r/2$ 
  as given in \eql{tb8}{b}.
  From \eqn{tb2}, \eqn{tb3} and \eqn{tb3b}, we find that
  $R=R^S\,R^L=R^L\,R^S$ and so, in the $J$RF,
  \bln{tb8c}
  R(\phi,\theta,0)\,|\,\ell\,m_\ell\;S\,m\ket
  =\sum_{m'\,m'_\ell}\,|\,\ell\,m'_\ell\;S\,m'\ket\,D^\ell_{m'_\ell\,m_\ell}(\phi,\theta,0)\,D^S_{m'\,m}(\phi,\theta,0)
  \eln
  which shows that the state $|\,\ell\,m_\ell\;S\,m\ket$ is a product of 
  two states $|\ell\,m_\ell\ket$ and $|S\,m\ket$ at rest in the $J$RF.
  We can thus derive the standard expansion of the relativistic two-body
  system in the $\ell S$-coupling scheme
  \bln{tb9}
  |JM\,\ell S\ket=\sum_{m_\ell\,m}\,(\ell\,m_\ell\;S\,m|JM)\,|\,\ell\,m_\ell\;S\,m\ket
  \eln
  or
  \bln{tb10}
  |JM\,\ell S\ket&\propto\sum_{m_\ell\,m}\,(\ell m_{\ell}\,Sm|JM)
  \sum_{m_s\,m_\sigma}\,(sm_s\;\sigma m_\sigma\,|\,S\,m)\cr
  &\hskip24mm\times\int{\rm d}\Omega\,Y^\ell_{m_\ell}(\Omega)|\,\vec q,\;s\,m_s\ket\;|\,\vec k,\;\sigma\,m_\sigma\ket
  \eln
  where $\vec q=\vec r/2$ and $\vec k=-\vec r/2$ in the $J$RF. We obtain 
  \bln{tb11}
  R(\phi,\theta,0)\,|JM\,\ell S\ket=\sum_{M'}\,|JM'\,\ell S\ket\,D^J_{M'\,M}(\phi,\theta,0)
  \eln
  This is the standard result, showing that $J$, $S$ and $\ell$ are
  rotational invariants.
  
  We may summarize the results of this section as follows: There exist two
  ket states,  $|S\,m\ket$ and $|\ell\,m_\ell\ket$,
  in the decay process $J\to s+\sigma$.
  They act like the states at rest in the $J$RF, as shown in \eql{tb8b}{a}
  and \eql{tb8b}{b}.
  So there exist three at-rest states in the problem; $|JM\ket$ in the initial
  system and $|\ell\,m_\ell\ket$ in the final system, at rest in the $J$RF.
  The third state $|S\,m\ket$ acts as if it is again at rest in the $J$RF,
  but it is composed of two states $|\,\vec r/2,\;s\,m_s\ket$ and 
  $|\,-\vec r/2,\;\sigma\,m_\sigma\ket$  ``in motion'' in the $J$RF.
  We emphasize here that this is true even when the parent mass becomes
  arbitraily large; the state in a total intrinsic spin $S$, as given by
  \eqn{tb7}, acts like a state at rest in the $J$RF. 
  We shall exploit this in Section 6 to derive our covariant decay amplitudes
  in the momentum space.
\sectn{Two-Body Decay $J\to s+\sigma$}{sc3}
\indnt
Consider a two-body decay $J\to s+\sigma$, where we use the notations very similar
to those introduced in a previous paper\cite{SCh0}. The daughter 1 with spin $s$ is described by
a wave function $\omega(\lambda)$ and the daughter 2 by $\varepsilon(-\nu)$\footnote{
The momentum of the daughter
$\sigma$ is pointed toward the negative $z$-axis, as shown by $k^\alpha$ in \eqn{pqk}.
By definition, its helicity $\nu$ is measured along $\vec k$; as a consequence,
its spin projection along the positive $z$-axis is $-\nu$.};
see Table I for all other relevant quantities.
\noindent\parbox[]{16cm}{
\begin{center}
Table I.\hspace{4mm} Two-body decay: $J\to s+\sigma$\hfill
\end{center}
\begin{center}
\begin{tabular}[]{|l|c|c|c|}\hline
                 & Parent     & Daughter 1  & Daughter 2\\\hline
Spin             & $J$        & $s$         & $\sigma$\\
Parity           & $\eta_{_J}$& $\eta_s$    & $\eta_\sigma$\\
Helicity         &            & $\lambda$   & $\nu$\\
Momentum         & $\vct{p}$        & $\vct{q}$         & $\vct{k}$\\
Energy           & $p_0$      & $q_0$       & $k_0$\\
Mass             & $w$        & $w_s$         & $w_\sigma$\\
Energy/Mass      &            & $\gamma_s$  & $\gamma_\sigma$ \\
Velocity         &            & $\beta_s$   & $\beta_\sigma$\\
Wave function&$\phi^*(\lambda-\nu)$&$\omega(\lambda)$&$\varepsilon(-\nu)$\\
\hline
\end{tabular}
\end{center}}

   The polarization four-vectors or wave functions 
appropriate for the particles $J=1$, `$s=1$' and `$\sigma=1$' are 
well known.  Along with the relevant momenta, 
\bln{pqk}\begin{tabular}{lrlrrrrl}
    $p^\alpha$ &= &(&      $w$;&0,&0,& 0&) \\
    $q^\alpha$ &= &(&    $q_0$;&0,&0,& $q$&)\\
	     &= &(&$\gamma_s w_s$;&0,&0,&$\gamma_s\beta_s w_s$&)\\
    $k^\alpha$ &= &(&    $k_0$;&0,&0,&$-q$&)\\
   &= &(&$\gamma_\sigma w_\sigma$;&0,&0,&$-\gamma_\sigma\beta_\sigma w_\sigma$&)\\
    $r^\alpha$ &= &(&$q_0-k_0$;&0,&0,&$2q$&)\\
\end{tabular}\eln\noindent
where $w=q_0+k_0$,\ $q_0=\sqrt{w_s^2+q^2}$,\ $k_0=\sqrt{w_\sigma^2+q^2}$
and $r=q-k$,
the wave functions in the $J\,$RF are given by
\bln{polz}\begin{tabular}{lrrrrrrrr}
   $\phi^\alpha(\pm)$  &=&$\mp\frac{1}{\sqrt{2}}$&(&0;&1,&$\pm i$,&0&)\\
   $\phi^\alpha(0)$    &=&                   &(&0;&0,&    0,&1&)\\
   $\omega^\alpha(\pm)$&=&$\mp\frac{1}{\sqrt{2}}$&(&0;&1,&$\pm i$,&0&)\\
   $\omega^\alpha(0)$  &=&& (&$\gamma_s\beta_s$;&0,&0,& $\gamma_s$&)\\
   $\varepsilon^\alpha(\pm)$  &=&$\mp\frac{1}{\sqrt{2}}$&(&0;&1,&$\pm i$,&0&)\\
   $\varepsilon^\alpha(0)$&=&& (&$-\gamma_\sigma\beta_\sigma$;
               &0,&0,& $\gamma_\sigma$&)\\
\end{tabular}\eln\noindent
Note that 
$$
p_\alpha\phi^\alpha(\lambda)=q_\alpha\omega^\alpha(\lambda)
   =k_\alpha\varepsilon^\alpha(\lambda)=0
$$
for any $\lambda$.

   These polarization four-vectors satisfy
\bln{Nrm1}
          p_\alpha\phi^\alpha(m)&=\ 0  \cr
        \phi^*_\alpha(m)\phi^\alpha(m')&=-\delta_{mm'}  \cr
  \sum_m \phi_\alpha(m)\phi^*_\beta(m)&=\ \tilde g_{\alpha\beta}(w)  
\eln
where
\bln{Lmt}
     \tilde g_{\alpha\beta}(w)
            =-g_{\alpha\beta}+\frac{p_\alpha p_\beta}{w^2}
\eln
The last equation of \eqn{Nrm1} 
is the usual projection operator for spin-1 states\cite{SCh0}.  Note that, in the
$J\,$RF, $\tilde g(w)$ has a zero time-component and +1 for the
space-components, i.e.  
\bln{gW}
   \tilde g_{\alpha\beta}(w)=\tilde g^{\alpha\beta}(w)=
\left(\begin{array}{ccccccc}
  0 &\phantom{00}& 0 &\phantom{00}& 0 &\phantom{00}& 0\\
  0 &\phantom{00}& 1 &\phantom{00}& 0 &\phantom{00}& 0\\
  0 &\phantom{00}& 0 &\phantom{00}& 1 &\phantom{00}& 0\\
  0 &\phantom{00}& 0 &\phantom{00}& 0 &\phantom{00}& 1
\end{array}\right)
\eln
$\omega$ and $\varepsilon$ satisfy similar conditions, 
but with their own $\tilde g$'s, i.e., $\tilde g(w_s)$ 
and $\tilde g(w_\sigma)$
\bln{g0}
     \tilde g_{\alpha\beta}(w_s)
            &=-g_{\alpha\beta}+\frac{q_\alpha q_\beta}{w_s^2}\cr
     \tilde g_{\alpha\beta}(w_\sigma)
            &=-g_{\alpha\beta}+\frac{k_\alpha k_\beta}{w_\sigma^2}\cr
\eln
so that
\bln{g1}
          q_\alpha\omega^\alpha(m)&=\ 0  \cr
        \omega^*_\alpha(m)\omega^\alpha(m')&=-\delta_{mm'}  \cr
  \sum_m \omega_\alpha(m)\omega^*_\beta(m)&=\ \tilde g_{\alpha\beta}(w_s)  
\eln
and
\bln{g2}
          k_\alpha\varepsilon^\alpha(m)&=\ 0  \cr
        \varepsilon^*_\alpha(m)\varepsilon^\alpha(m')&=-\delta_{mm'}  \cr
  \sum_m \varepsilon_\alpha(m)\varepsilon^*_\beta(m)&=\ \tilde g_{\alpha\beta}(w_\sigma)  
\eln
\sectn{Spin-$J$ Wave Functions}{sc4}
\indnt
   The general spin-$J$ wave function can be written\cite{SCh0}, with $m\geq0$
\bln{s0}
  \phi_{\delta_1\cdots\delta_J}(J\,m)
  &=[a^J(m)]^\frac{1}{2}\sum_{m_0}\,2^{(m_0-\kappa)/2}\,\sum_P\,
  \underbrace{\phi_{\alpha_1}(+)\cdots}_{m_+}
	\underbrace{\phi_{\beta_1}(0)\cdots}_{m_0}
	\underbrace{\phi_{\gamma_1}(-)\cdots}_{m_-}\cr
  &{\rm with}\quad a^J(m)=\frac{2^\kappa\,(J+m)!(J-m)!}{(2J)!}
\eln
The indices $\{\delta_1\cdots\delta_J\}$
are grouped into
$\{\alpha_i\}$ with $(i=1,m_+)$, $\{\beta_i\}$ with $(i=1,m_0)$ 
and $\{\gamma_i\}$ with $(i=1,m_-)$, where $m_\pm$ stands for the number
of $\phi(\pm)$'s and $m_0$ for the number of $\phi(0)$'s.
Here $\phi_{\delta_j}(m)$ is the spin-1 wave function
introduced in Section 3.  The second sum above stands for the sum of
all distinct permutations of the $\phi_{\delta_j}$'s.

The possible $m_\pm$ and $m_0$ are restricted by
\bln{s1}
   J=m_++m_0+m_-,\quad m=m_+-m_-,\quad m_+\geq m_-
\eln
such that
\bln{s2}
   2m_\pm=J\pm m-m_0
\eln
$m_0$ ranges from $0(1)$, $2(3)$,$\cdots$, to $J-m=$even(odd). Since 
$\phi_{\delta_1\cdots\delta_J}$ does not depend on $\kappa$, one can choose
$\kappa=0(1)$ for $J-m=$even(odd), such that the coefficients under the sum
on $m_0$
do {\em not} have square-root factors in them.

   The wave functions for $J=2$ are
\bln{s3}
\phi_{\alpha\beta} (22) &=\phi_\alpha(+)\, \phi_\beta(+)\cr
\sqrt{2}\,\phi_{\alpha\beta} (21) &=
\phi_\alpha(+)\, \phi_\beta(0)+\phi_\alpha(0)\, \phi_\beta(+)\cr
\sqrt{6}\,\phi_{\alpha\beta} (20) &=
\phi_\alpha(+)\, \phi_\beta(-)+\phi_\alpha(-)\, \phi_\beta(+)
	+2\phi_\alpha(0)\, \phi_\beta(0)
\eln
and, for $J=3$,
\bln{s4}
 \phi_{\alpha\beta\gamma}(33)&=
          \phi_\alpha(+)\,\phi_\beta(+)\,\phi_\gamma(+)\cr
\sqrt{3}\, \phi_{\alpha\beta\gamma}(32)&=\phi_\alpha(+)\,\phi_\beta(+)\,\phi_\gamma(0)
    +\phi_\alpha(+)\,\phi_\beta(0)\,\phi_\gamma(+)
       +\phi_\alpha(0)\,\phi_\beta(+)\,\phi_\gamma(+)\cr
\sqrt{15}\, \phi_{\alpha\beta\gamma}(31)&=\phi_\alpha(+)\,\phi_\beta(+)\,\phi_\gamma(-)
   +\phi_\alpha(+)\,\phi_\beta(-)\,\phi_\gamma(+)
       +\phi_\alpha(-)\,\phi_\beta(+)\,\phi_\gamma(+)\cr
      &\hskip-12mm+{2}\Big[\phi_\alpha(+)\,\phi_\beta(0)\,\phi_\gamma(0)
    +\phi_\alpha(0)\,\phi_\beta(+)\,\phi_\gamma(0)
       +\phi_\alpha(0)\,\phi_\beta(0)\,\phi_\gamma(+)\Big]\cr
\sqrt{10}\, \phi_{\alpha\beta\gamma}(30)&=\phi_\alpha(0)\,\phi_\beta(+)\,\phi_\gamma(-)
    +\phi_\alpha(0)\,\phi_\beta(-)\,\phi_\gamma(+)
       +\phi_\alpha(-)\,\phi_\beta(+)\,\phi_\gamma(0)\cr
         &\hskip-12mm+\phi_\alpha(+)\,\phi_\beta(-)\,\phi_\gamma(0)
    +\phi_\alpha(+)\,\phi_\beta(0)\,\phi_\gamma(-)
       +\phi_\alpha(-)\,\phi_\beta(0)\,\phi_\gamma(+)\cr
           &\hskip24mm+2\phi_\alpha(0)\,\phi_\beta(0)\,\phi_\gamma(0)\cr
\eln
For the case $m<0$, the wave functions can be obtained by using
\bln{s2a}
   \phi(J\,-\!m)=(-)^m\;\phi^*(J\,m)
\eln

   In the $J\,$RF, the wave functions $\phi(J\,m)$ have a simple form, i.e.
the time components are zero as shown in the previous section for $J=1$.
The wave functions for $s=1$ and $\sigma=1$ have been given in \eqn{polz}.
So the wave functions for arbitrary spins $s$ and $\sigma$ are given again by \eqn{s0},
in which the spin-1 wave functions $\phi_\alpha(m)$'s within the sum $\sum_P$
have been replaced by $\omega_\alpha(m)$'s and $\varepsilon_\alpha(m)$'s, respectively.
Note that the general wave functions $\omega_\alpha(s\,m)$ and $\varepsilon_\alpha(\sigma\,m)$,
where $s$ and $\sigma$ are arbitrary integer, have nonzero components for all of their
Lorentz indices 0, 1, 2 and 3.

We now turn to the treatment of orbital-angular momenta in our formulation. 
We have shown in \eql{tb8b}{b} that the state $|\ell\,m_\ell\ket$ which describes 
the orbital-angular momentum $\ell$ acts like a state at rest in the $J$RF.  It is clear, therefore, 
that the correct wave function for $\ell$ must be described by a symmetric and traceless tensor 
of rank $\ell$ whose space components correspond to the relative momentum $\vec r$ 
but with {\em zero} time-component in the $J$RF.  
There are, in fact, two wave functions with zero time components in the $J$RF;
a rank-$J$ tensor $\phi^*(Jm)$ in the initial state and one $\chi(\ell\,0)\,r^\ell$
representing the orbital angular momentum $\ell$ in the final state.  They obey, be definition,
$p^\mu\,\phi^*_\mu(Jm)=0$ and $p^\mu\,\chi_\mu(\ell\,0)=0$.
Since we deal exclusively with collinear vectors aligned along the $z$-axis, the spin projection of
the wave function comes with a zero $z$-component only, reflecting the fact that orbital angular
momentum must be perpendicular to the $z$-axis [see \eql{tb3b}{b}].  So we are left with the task of evaluating,
in the $J\,$RF, a symmetric and traceless tensor\cite{SCh0}
\bln{s6}
   \chi_{ijk\cdots}(\ell0)=\frac{(\ell!)^2}{(2\ell)!}\sum_{m_0}2^{(\ell+m_0)/2}
	\sum_P\,\bigg[\underbrace{\chi(+)\cdots}_{m_+}
		\underbrace{\chi(0)\cdots}_{m_0}
		\underbrace{\chi(-)\cdots}_{m_-}\bigg]_{ijk}
\eln
and $m_0=0(1),\ 2(3),\ 4(5),\ldots$ for $\ell=$even (odd).  It is clear that
$\chi(\ell0)$ is devoid of the coefficients with square-root factors in the sum.
We find
\bln{s7}
	\chi(0)&=(0,0,1),\quad\chi(\pm)=\mp \frac{1}{\sqrt{2}} \,(1,\pm i,0)\cr
  3\,\chi_{ij}(20)&=\chi_i(+)\,\chi_j(-)+\chi_i(-)\,\chi_j(+)
		+2\,\chi_i(0)\,\chi_j(0)\cr
  5\,\chi_{ijk}(30)&=\chi_i(0)\,\chi_j(+)\,\chi_k(-)+\chi_i(0)\,\chi_j(-)\chi_k(+)\cr
	&+\chi_i(+)\,\chi_j(0)\,\chi_j(-)+\chi_i(-)\,\chi_j(0)\chi_k(+)\cr
	&+\chi_i(+)\,\chi_j(-)\,\chi_j(0)+\chi_i(-)\,\chi_j(+)\chi_k(0)
		+2\,\chi_i(0)\,\chi_j(0)\,\chi_k(0)\cr
\eln
and
\bln{s8}
  35\,\chi_{ijkl}(40)
	&=2\sum_P^6\bigg[\chi_i(+)\,\chi_j(-)\,\chi_k(+)\,\chi_l(-)\bigg]\cr	
	&\hskip-12mm+4\sum_P^{12}\bigg[\chi_i(0)\,\chi_j(0)\,\chi_k(+)\,\chi_l(-)\bigg]
		+8\,\chi_i(0)\,\chi_j(0)\,\chi_k(0)\,\chi_l(0)\cr
  63\,\chi_{ijklm}(50)
&=2\sum_P^{30}\bigg[\chi_i(0)\,\chi_j(+)\,\chi_k(-)\,\chi_l(+)\,\chi_m(-)\bigg]\cr	
&\hskip-12mm+4\sum_P^{20}\big[\chi_i(0)\,\chi_j(0)\,\chi_k(0)\,\chi_l(+)\,\chi_m(-)\big]
		+8\,\chi_i(0)\,\chi_j(0)\,\chi_k(0)\,\chi_l(0)\,\chi_m(0)\cr
\eln
 
   It should be noted that $\chi(\ell\,0)$ is given only in space indices 1, 2 and 3.
The orbital angular momentum in the $J\,$RF does not have time components, and therefore
it is in fact proportional\cite{SCh0} to $\phi(J\,m)$ in the $J\,$RF with $J=\ell$ and $m=0$
\bln{s9}
   \chi_{ijk\cdots}(\ell0)=c_\ell\;\phi_{ijk\cdots}(\ell0),\quad
	c_\ell=\left[ \frac{2^\ell\,(\ell!)^2}{(2\ell)!} \right]^{1/2}
\eln
The coefficients $c_\ell$ for a few values of $\ell$ are
\bln{s10}
   c_1=1,\quad c_2=\sqrt{\frac{2}{3}},\quad c_3=\sqrt{\frac{2}{5}},\quad c_4=\sqrt{\frac{8}{35}}
\eln
\sectn{Wave Functions of Total Intrinsic Spin $S$}{sc5}
\indnt
   The wave function $\psi(s,\sigma;S\delta)$ corresponds
to the state of total intrinsic spin $S$, built out of the wave functions for
$s$ and $\sigma$ which are given by the tensors of rank-$s$ and -$\sigma$, respectively.
Hence, it is a tensor of rank $s+\sigma$, given by
\bln{a1a0}
\psi(s,\sigma;S\delta)&=\sum_{m_a\,m_b}\,(sm_a\;\sigma m_b|S\delta)\,
\omega(sm_a)\,\varepsilon(\sigma m_b)
\eln 
where we have suppressed tensor indices in $\psi$, $\omega$ and $\varepsilon$ [see \eqn{tb7}].
In addition, we have suppressed the momentum labels in the wave functions.

   Consider now the special case $s=\sigma=1$.  
Then, the $\psi(s,\sigma;S\delta)$ is a rank-2 tensor.
In the $J\,$RF, a state in total intrinsic spin $S$ is
\bln{a1a}
   \psi^{\mu\nu}(s,\sigma;Sm_{_S})
	&=\sum_{m_1m_2}\,(sm_1\,\sigma m_2|Sm_{_S})\,\omega^\mu(m_1)\,\varepsilon^\nu(m_2)\cr
   \psi^{\alpha\beta}(s,\sigma;S\,-\!m_{_S})
	&=(-)^{S-m_s}\,\psi^{\alpha\beta\cnj}(s,\sigma;Sm_{_S})
\eln
Under an arbitrary rotation $R^S(\phi,\theta,0)$, the daughter states
$s$ and $\sigma$ transform
\bln{a1b}
R^S(\phi,\theta,0)\,\omega^\mu(m_1)&\equiv
	{\bigl[R^S(\phi,\theta,0)\bigr]^\mu}_\alpha\;\omega^\alpha(m_1)
	=\sum_{m'_1}\,\omega^{\mu}(m'_1)\,
	D^s_{m'_1\,m^{\ftm{'}}_1}(\phi,\theta,0)\cr
R^S(\phi,\theta,0)\,\varepsilon^\nu(m_2)&\equiv
{\bigl[R^S(\phi,\theta,0)\bigr]^\nu}_\beta\;\varepsilon^\beta(m_2)
	=\sum_{m'_2}\,\omega^{\nu}(m'_2)\,
	D^\sigma_{m'_2\,m^{\ftm{'}}_2}(\phi,\theta,0)\cr
\eln
so that
\bln{a1c}
   R^S(\phi,\theta,0)\,\psi^{\mu\nu}(s,\sigma;Sm_{_S})
	=\sum_{m'_s}\,\psi^{\mu\nu}(s,\sigma;Sm'_s)\,
	D^S_{m'_s\,m^{\ftm{'}}_s}(\phi,\theta,0)
\eln
This shows in fact that $\psi^{\mu\nu}(s,\sigma;Sm_{_S})$ 
is in a state of total intrinsic spin $S$.
It is useful to write down the wave functions explicitly
\blb{c1d}
\psi^{\alpha\beta}(s,\sigma;22)&=\omega^\alpha(+)\,\varepsilon^\beta(+),\quad\lambda=-\nu=+1\cr  
\psi^{\alpha\beta}(s,\sigma;21)&=\frac{1}{\sqrt{2}} \biggl[\omega^\alpha(+)\,\varepsilon^\beta(0)
	+\omega^\alpha(0)\,\varepsilon^\beta(+)\biggr]\cr  
\psi^{\alpha\beta}(s,\sigma;20)&=\frac{1}{\sqrt{6}} \biggl[\omega^\alpha(+)\,\varepsilon^\beta(-)
   +\omega^\alpha(-)\,\varepsilon^\beta(+)+2\omega^\alpha(0)\,\varepsilon^\beta(0)\biggr]\cr  
\psi^{\alpha\beta}(s,\sigma;11)&=\frac{1}{\sqrt{2}} \biggl[\omega^\alpha(+)\,\varepsilon^\beta(0)
	-\omega^\alpha(0)\,\varepsilon^\beta(+)\biggr]\cr  
\psi^{\alpha\beta}(s,\sigma;10)&=\frac{1}{\sqrt{2}} \biggl[\omega^\alpha(+)\,\varepsilon^\beta(-)
   -\omega^\alpha(-)\,\varepsilon^\beta(+)\biggr]\cr  
\psi^{\alpha\beta}(s,\sigma;00)&=\frac{1}{\sqrt{3}} \biggl[\omega^\alpha(+)\,\varepsilon^\beta(-)
   +\omega^\alpha(-)\,\varepsilon^\beta(+)-\omega^\alpha(0)\,\varepsilon^\beta(0)\biggr]\cr  
\elb
In applications involving $\psi^{\mu\nu}(s,\sigma;Sm_{_S})$ listed above, we can safely ignore
the non-zero time components in the wave functions $\omega^\alpha(m_s)$ and
$\varepsilon^\beta(m_\sigma)$, because they are generally mated with $\chi(\ell\,0)$ and
$\phi^*(m)$ which have {\em no} time components in the $J\,$RF.  The sole exception
occurs in the treatment of the decay $0^+\to 1^-+1^-$, where the non-zero time
components of $\psi^{\mu\nu}(s,\sigma;Sm_{_S})$ could {\em possibly} enter; this case is
considered in detail in the next section.  See also Appendix C for additional decay amplitudes 
which involve the time components of $\psi^{\mu\nu}(s,\sigma;Sm_{_S})$  but not
included in the main text.
\sectn{Covariant Helicity-Coupling Amplitudes}{sc}
\indnt
   The amplitude for $J\to s+\sigma$ in the $J\,$RF is, in the helicity formalism\cite{SCh2},
\bln{a0}
   {\cal M}^J_{\lambda\nu}(M;\theta,\phi)=\sqrt{\frac{2J+1}{4\pi}}\;F^J_{\lambda\,\nu}\,
	D^{J\cnj}_{M\,\delta}(\phi,\theta,0),\quad\delta=\lambda-\nu
\eln
where $F^J_{\lambda\,\nu}$ is the helicity-coupling amplitude given by
\bln{a1}
   F^J_{\lambda\nu}=\sum_{\ell S}\,\left(\frac{2\ell+1}{2J+1}\right)^{1/2}\!\!
	(\ell0\,S\delta|J\delta)\;(s\lambda\;\sigma\,-\!\nu|S\delta)\;G^{J}_{\ell S} \,r^\ell\cr
\eln
Here $G^{J}_{\ell S}$ is the $\ell S$-coupling amplitude for which $\ell$ is the
orbital angular momentum and $S$ is the total intrinsic spin, and  
$r^\ell$ is the barrier factor.  The variable $r$  (see Table I)
is measured in units of the scale factor $r_0$(e.g. 1\,fm),
so should be understood as (unit-free) quantity
$(r/r_0)^\ell$.
For further comments on this point, 
see Conclusions and Discussions.  
Note that the rotationally invariant quantum numbers are given as super- and sub-scripts
in the amplitude ${\cal M}^J_{\lambda\nu}$, $F^J_{\lambda\nu}$ and $G^{J}_{\ell S}$.  
Note that we must in principle include the spins $s$ and $\sigma$
as well in the amplitudes, i.e. the superscript $J$ should be replaced by $\{J\,s\,\sigma\}$.
However, we shall continue to use simply the superscript $J$ for brevity of notation.
The number of independent $F^J_{\lambda\nu}$'s is the same
as that of $G^{J}_{\ell S}$, as each of them describes the degrees of freedom 
for the decay process in the helicity and the canonical formalisms.  For a general description
of the number of independent $F^J_{\lambda\nu}$'s, see ref.~\cite{SCh0}.

   The coefficients of $G^{J}_{\ell S} \,r^\ell$ in \eqn{a1} satisfy
\bqx{a1x0}
  &\sum_{\ell\, S}\,\left(\frac{2\ell+1}{2J+1}\right)
   (\ell0\,S\delta|J\delta)^2\;(s\lambda\;\sigma\,-\!\nu|S\delta)^2=1
	\quad\mbox{for a given}\ \{\lambda,\nu\}&\mylbl{a}\cr
  &\sum_{\lambda\,\nu}\,\left(\frac{2\ell+1}{2J+1}\right)^{1/2}\hskip-8pt
   (\ell0\,S\delta|J\delta)\;(s\lambda\;\sigma\,-\!\nu|S\delta)
	\left(\frac{2\ell'+1}{2J+1}\right)^{1/2}\hskip-8pt
   (\ell'0\,S'\delta|J\delta)\;(s\lambda\;\sigma\,-\!\nu|S'\delta)\cr
   &\hskip60mm
  	=\delta_{\,\ell\,\ell'}\;\delta_{\,S\,S'}  
		\quad\mbox{for a given}\ \{\,\ell, S,\ell', S'\,\}&\mylbl{b}
\eqx
with the proviso that $\ell$ and $\ell'$ can be either always even or always odd, 
from parity conservation in the decay $J\to s+\sigma$.
The relationship \eql{a1x0}{a} shows that the sum over a set $\{\lambda,\nu\}$ of the squares 
of all the coefficients of a given $G^{J}_{\ell S} \,r^\ell$ in \eqn{a1} should be equal to one.
The  relationship \eql{a1x0}{b} for $\ell'=\ell$ and $S'=S$ can be used to show that
the sum over the squares of the coefficients of $G^{J}_{\ell S} \,r^\ell$
in a given $F^J_{\lambda\nu}$ in \eqn{a1} must be equal to one.  
We will show later with a number of simple examples how these constraints are borne out.
From \eql{a1x0}{b} we obtain
\bln{a1x1}
  \sum_{\lambda\,\nu}\Big|F^J_{\lambda\nu}\Big|^2
	=\sum_{\ell\,S}\Big|G^{J}_{\ell S} \,r^\ell\,\Big|^2
\eln
This simply shows that probability is conserved for both canonical and helicity 
description for the decay process $J\to s+\sigma$.

   Consider now $\theta=\phi=0$.  The $D$-function in \eqn{a0} is zero unless $M=\delta$.
Hence we can write
\bln{a2}
   F^J_{\lambda\,\nu}=\sqrt{\frac{4\pi}{2J+1}}\;{\cal M}^J_{\lambda\,\nu}(\delta;0,0)
\eln
So the helicity-coupling amplitude is proportional to the decay amplitude itself
in which the the decay products $s$ and $\sigma$ are both aligned along the $z$-axis,
in a coordinate system fixed in the $J\,$RF.  Hence, the momenta $\vct{q}$ and $\vct{k}$
have zero $x$ and $y$ components; see \eqn{pqk}.  

   We now transform the helicity-coupling amplitude \eqn{a1} by replacing,
formally, the factor
\bln{a2a}
	\left(\frac{2\ell+1}{2J+1}\right)^{1/2}\!\!(\ell0\,S\delta|J\delta)\;G^{J}_{\ell S} \,r^\ell\cr
\eln
with that proportional to a decay amplitude and write
\bln{a10}
   F^J_{\lambda\nu}&=\sum_{\ell S}
(s\lambda\;\sigma\,-\!\nu|S\delta)\;g^J_{\ell S}\,\,A^J_{\ell S}(s,\sigma;\delta)\,r^\ell\cr
\eln
We assume that
$g^J_{\ell S}$ is a constant to be measured experimentally, 
while $A^J_{\ell S}(s,\sigma;\delta)$
is a {\em covariant} decay amplitude  corresponding to the transition
$|J\delta\ket\to|S\delta\ket+|\ell0\ket$
in which the decay products are {\em aligned} along the $z$-axis.
The precise relationship between $g^J_{\ell S}$ and $G^{J}_{\ell S}$
will be given at the end of this section. 
It turns out that in general,
there is more than one relativistic structure for a given \{$\ell S$\}.
This is a consequence of treating $\gamma_s$ and $\gamma_{\sigma}$ as free
parameters, which appear as factors in the relativistic terms
related to a non-relativistic $G^J_{\ell S}$.
This dependency on the particle masses can be disentangled if the
widths of the daughter particles are sufficiently wide and  the distribution 
function is examined over a sufficiently wide range of the parent mass.

The procedure for constructing covariant amplitudes goes as follows.
To guarantee Lorentz invariance, one utilizes
two {\em invariant tensors} of the Poincar\'e group\cite{MG}, i.e.
\bln{a10a}
   g^{\mu\nu}&=g^{\alpha\beta}\,{\Lambda_{\alpha}}^\mu\,{\Lambda_{\beta}}^\nu\cr
  \varepsilon^{\,\nu\,\rho\,\sigma\,\tau}
    &=\varepsilon^{\,\mu\,\alpha\,\beta\,\gamma}\,\Lambda_\mu{}^\nu\,
	\Lambda_\alpha{}^\rho\,\Lambda_\beta{}^\sigma\,
	\Lambda_\gamma{}^\tau
\eln
where ${\Lambda_{\alpha}}^\mu$ is an element of the homogeneous Lorentz group (see also Appendix A).
Now,
we adopt a more restrictive procedure to ensure that the resulting amplitudes
have a transparent transition to non-relativistic amplitudes.  For the purpose,
we write, in the $J\,$RF, Lorentz-invariant decay amplitudes in the $\ell S$ coupling scheme via
\bln{a11}
   A^J_{\ell S}(s,\sigma;\delta)
	  &=\big[\,p^{n_0},\; \psi(s,\sigma;S\delta),\;
			\chi(\ell0),\;\phi^*(J\delta)\,\big]_w\cr
\eln
where $n_0=0$ or 1 and the notation $[\cdots]_w$ stands for Lorentz contraction 
of all the relevant Lorentz indices in the $J\,$RF.
We adopt the following two rules:
\begin{list}{$\bullet$}
	{\topsep=-2pt\itemsep=-2pt\leftmargin=36pt\rightmargin=36pt}
\item[(a)]with the modified metric $\tilde g(w)$ [\,see \eqn{Lmt}\,] 
and {\em only} with this metric, if $s+\sigma+\ell-J=$ even.  Here $p^\mu$ never appears
in the amplitudes, i.e. $n_0=0$.
\item[(b)]with the totally antisymmetric rank-4 tensor (\,$\epsilon^{\alpha\beta\mu\nu}$\,) 
{\em together with} the 4-momentum $p^\mu$, 
\undl{and} with $\tilde g(w)$ whenever necessary, if $s+\sigma+\ell-J=$ odd.  
We restrict the use of $\epsilon^{\alpha\beta\mu\nu}$ and $p^\mu$ once and only once,
so that $n_0=1$.
\item[(c)] For angular momentum $\ell$, only amplitudes proportional to 
$r^\ell$ are considered ($r^\ell$-rule).
\end{list}
It is seen that $p^\mu=(w;0,0,0)$ is used in two different ways, once inside the
definition of $\tilde g(w)$, and again with the rank-4 tensor if necessary.
Since collinear vectors are used in $A^J_{\ell S}(s,\sigma;\delta)$,  
it cannot give angular dependence; but instead it gives
dependence on the Lorentz factors $\gamma_s$ and $\gamma_\sigma$---which is the reason for
calculating the invariant amplitudes in this manner.  The decay amplitudes which lie beyond
the scope of the rules above, are treated in Appendix C, and the reasons
explained for not including in the examples given in this paper.

   It will be shown later that the presence of the rank-4 tensor along with $p$ 
in the amplitude \eqn{a11} leads to a factor $(i\,w)$ when evaluated in the $J\,$RF.
As the wave functions which appear in \eqn{a11} are all unitless,  we can impose the unitless
condition in \eqn{a11} by substiuting $p^{n_0}$  by $\big[p/(i\,w_{\rm th})\big]^{n_0}$ 
where $w_{\rm th}$ is the $w$ at its threshold.  So the factor $(i\,w)$ which appears the
amplitudes is then replaced by $(w/w_{\rm th})$. However, we will continue to use $p^{n_0}$
and the factor $(i\,w)$ in this paper for simplicity of notation.
The reader will note that the situation here is similar to that for $r^\ell$ against
$(r/r_0)^\ell$ commented on previously in this section.  It is perhaps more appropriate
to write the covariant amplitudes as 
\bln{a11a}
   A^J_{\ell S}(s,\sigma;\delta)=
      \bigg[\,\left(\frac{p}{i\,w_{\rm th}}\right)^{n_0},\; \psi(s,\sigma;S\delta),\;
			\chi(\ell0),\;\phi^*(J\delta)\,\bigg]_w
\eln
Note that the right-hand side is now explicitly unitless.  We will come back to this point again
in Conclusions and Discussions.

   The wave function $\phi^*(J\delta)$ refers to the initial spin $J$ 
and $\chi(\ell0)$ to
the orbital angular momentum $\ell$ in the final state.  They are
rank-$J$ and rank-$\ell$ tensors, respectively. 
The wave functions of total intrinsic spin $S$, denoted by $\psi(s,\sigma;S\delta)$ in \eqn{a11},
has been given in \eqn{a1a0}.  It is a tensor of rank $s+\sigma$ built out of 
$\omega$'s and $\varepsilon$'s.
For the different possible Lorentz contractions, it is relevant how many
indices of the $\omega$'s and $\varepsilon$'s are contracted with $\phi$ and $\chi$.
For the purpose, it is necessary to introduce a new notation, 
in order to uniquely specify the process of Lorentz contraction.  
We transform $\psi$ to $\Psi$ as follows:
\bln{a11w1}
   	\psi(s,\sigma;Sm_{_S})\quad\longrightarrow\quad\Psi(n_1, n_2;Sm_{_S})
\eln 
where 
\bln{a11w2}
   n_1=n_{\omega\phi},\quad n'_1=n_{\omega\phi}+1,\qquad
   n_2=n_{\varepsilon\phi},\quad n'_2=n_{\varepsilon\phi}+1
\eln
Here $n_{\omega\phi}$ ($n_{\varepsilon\phi}$) is the number of
contractions of $\phi$  with $\omega$ ($\varepsilon$).  The reason for a change of notation
from $\psi$ to $\Psi$ is to merely indicate that the variables $\{s,\sigma\}$ of $\psi$
have been changed to $\{n_1,n_2\}$.
The meaning of apostrophes is explained shortly.
The remaining indices of $\omega$ and $\varepsilon$ are contracted with $\chi$ or
among themselves, except that, if necessary, 
three indices from  $\omega$, $\varepsilon$, $\chi$ and $\phi$ (one and only one index
from each wave function for a total of three), along with $p^\mu$,
are contracted with the totally antisymmetric rank-4 tensor
$\epsilon^{\alpha\beta\mu\nu}$.
This procedure is illustrated in Tables II through IV;
The heading for the column representing $\Psi(n_1, n_2;Sm_{_S})$ is
simply written as $R(n_1)$ if $s\geq1$ and $\sigma=0$ or $R(n_1,n_2)$ if $s\geq1$ and $\sigma\geq1$, 
where $R=s+\sigma$ is the rank of the tensor $\psi$.
One or two dots above $R$, i.e. $\onedot R$ or $\twodots R$, indicate 
one or two internal contractions between $\omega$ and $\varepsilon$.  
The number of dots between the the $\psi$, $\chi$ and $\phi$ indicate the number of indices
to be contracted between neighbors.  
If the contraction involving $\epsilon^{\alpha\beta\mu\nu}$ appears,
the appropriate indices are indicated by apostrophes (there are always three 
and only three apostrophes for each amplitude).

The helicity-coupling amplitudes satisfy, from parity conservation in the
strong decay,
\bln{a11x0}
F^J_{\lambda\nu}&=\eta_{_J}\,\eta_s\,\eta_\sigma\,(-)^{J-s-\sigma}\,F^J_{-\!\lambda\,-\!\nu}
=(-)^{s+\sigma+\ell-J}\,F^J_{-\!\lambda\,-\!\nu}
\eln
where the product of intrinsic parities is linked to the allowed orbital
angular momenta via $\eta_{_J}\,\eta_s\,\eta_\sigma=(-)^\ell$, so for a
given decay channel either $F^J_{\lambda\nu}=F^J_{-\!\lambda\,-\!\nu}$ or
$F^J_{\lambda\nu}=-F^J_{-\!\lambda\,-\!\nu}$ holds for all amplitudes.
From \eqn{a10}, we see that
\bln{a11x1}
A^J_{\ell S}(s,\sigma;\delta)=(-)^{\,\ell+S-J}\,A^J_{\ell S}(s,\sigma;-\delta)
\eln
If particles $s$ and $\sigma$ are identical, we have\cite{JW,SCh2}
\bln{a11x2}
   F^J_{\lambda\nu}(s,\sigma)=(-)^J\,F^J_{\nu\lambda}(\sigma,s)
\eln
and this leads to, from \eqn{a10},
\bln{a11x3}
   A^J_{\ell S}(s,\sigma;\delta)=(-)^J\,A^J_{\ell S}(\sigma,s;-\delta)
\eln
We impose the identical-particle condition by substituting
\bln{a11x4}
   F^J_{\lambda\nu}(s,\sigma)&\to
   \frac{1}{2}\left[F^J_{\lambda\nu}(s,\sigma)+(-)^J\,F^J_{\nu\lambda}(\sigma,s)\right]\cr
   A^J_{\ell S}(s,\sigma;\delta)&\to
   \frac{1}{2}\left[A^J_{\ell S}(s,\sigma;\delta)+(-)^J\,A^J_{\ell S}(\sigma,s;-\delta)\right]\cr
\eln
In this operation, we need to point out 
that the factor $r^\ell$ in \eqn{a10} must remain invariant;
in another words, `$r$' stands for the {\em magnitude} of `r' and hence it does not change sign.

   The formulas \eqn{a10} and \eqn{a11} constitute the {\em main result} of this paper.
We give a few examples of practical importance to illustrate the use of these formulas.
For the purpose, the following general formulas will prove to be useful.
Observe
\bln{a12}
   \phi^*(J\,m)&=(-)^m\,\phi(J\,-\!m)\cr	
   \omega^*(s\,m_s)&=(-)^{m_s}\,\omega(s\,-\!m_s),\quad
   \varepsilon^*(\sigma\,m_\sigma)=(-)^{m_\sigma}\,\varepsilon(\sigma\,-\!m_\sigma)\cr
	\chi^*(\ell\,0)&=\chi(\ell\,0)\cr
\eln
so we find
\bln{a13}
   \psi(s,\sigma;S\,-\!\delta)&=(-)^{s+\sigma-S}\,(-)^{\delta}\,\psi^*(s,\sigma;S\,\delta)\cr	
\eln
and
\bln{a14}
   A^J_{\ell S}(s,\sigma;-\delta)&=[p,\;\psi(s,\sigma;S\,-\!\delta),\;\chi(\ell0),\;
	\phi^*(J\,-\!\delta)]_w\cr
	&=(-)^{\,s+\sigma-S}\,A^{J\cnj}_{\ell S}(s,\sigma;\delta)\cr
\eln
and so, combining with \eqn{a11x1},
\bln{a141}
   A^{J\cnj}_{\ell S}(s,\sigma;\delta)=(-)^{\,s+\sigma+\ell-J}\,A^J_{\ell S}(s,\sigma;\delta)
\eln
So we see that $A^J_{\ell S}(s,\sigma;\delta)$ is {\em purely} real (imaginary) if
$s+\sigma+\ell-J$=even (odd).  We see that this is consistent with \eqn{a11}:
the amplitudes $A^J_{\ell S}(s,\sigma;\delta)$
are real in general, except when the $\epsilon^{\alpha\beta\gamma\delta}$ 
and $p$ are used to form a Lorentz scalar,
in which case the amplitudes are proportional to $(i\,w)$ times a real factor,
so that they become purely imaginary.
Since $(\ell0\,S\delta|J\delta)$ is contained in $A^J_{\ell S}(s,\sigma;\delta)$, we must have
\bln{a14a}
   A^J_{\ell S}(s,\sigma;0)=0,\qquad{\rm if}\quad\ell+S-J={\rm odd}
\eln
If there are more than one amplitude for a given $\{\ell\,S\}$, then we need to
include them all by modifying the appropriate factor in \eqn{a10}
\bln{a14b}
   g^J_{\ell S}\;A^J_{\ell S}(s,\sigma;\delta)
	\;\Rightarrow\;\sum_i\,^{(i)\!}g^J_{\ell S}\;\,^{(i)\!}A^J_{\ell S}(s,\sigma;\delta)
\eln
where $^{(i)\!}g^J_{\ell S}$'s are now the phenomenological constants 
to be determined experimentally.  In the examples given in this paper, there
are no more than three different amplitudes for a given $\{\ell\,S\}$;
so for a more transparent notation we denote the amplitudes $A$, $B$ and $C$,
with the constants $g$, $f$ and $h$, respectively.

   Let $\xi(m)$, $\xi'(m)$ and $\xi''(m)$ stand for any of the wave functions 
$\omega$, $\varepsilon$, $\chi$ or $\phi$. For the calculation of the
amplitudes it proves useful to note that
\blb{a12a}
   \vct{\xi}^*(\pm)&=-\vct{\xi}(\mp),\quad\vct{\xi}^*(0)=\vct{\xi}(0)\cr
 \vct{\xi}^*(m)\cdot\vct{\xi'}(m')&=\delta_{mm'},\quad{\rm if}\ m\ {\rm or}\ m' =\pm1,\cr
 \vct{\xi}^*(0)\cdot\vct{\xi'}(0)&=1, \quad {\rm if}\ 
    \xi\ {\rm and}\ \xi'\ =\ \chi\ {\rm or}\ \phi\cr
 \vct{\xi}^*(0)\cdot\vct{\xi'}(0)&=\gamma_s, \quad {\rm if}\ 
    \xi\ {\rm or}\ \xi' = \omega\ \mbox{and the other}=\chi\ {\rm or}\ \phi\cr
 \vct{\xi}^*(0)\cdot\vct{\xi'}(0)&=\gamma_\sigma, \quad {\rm if}\ 
    \xi\ {\rm or}\ \xi'=\varepsilon\ \mbox{and the other}=\chi\ {\rm or}\ \phi\cr
	\omega^*(0)\cdot\varepsilon(0)&=
	\omega(0)\cdot\varepsilon^*(0)=\omega(0)\cdot\varepsilon(0)=\gamma_s\gamma_\sigma
\elb
and also, using a short-hand notation where $\xi,\xi',\xi^{''}$
follow the sequence $\omega,\varepsilon,\chi,\phi$,
\bln{12b1}
[(m)(m')(m'')]&=[p,\xi(m),\xi'(m'),\xi^{''\cnj}(m'')]=0,\cr
\eln
except the following:
\blb{12b2}
[(\pm)(\mp)(0)]&=\pm(i\,w)a\cr
[(\pm)(0)(\pm)]&=\pm(i\,w)a\cr
[(0)(\pm)(\pm)]&=\mp(i\,w)a\cr
\elb
where
\blb{a12c}
 a&=1,\qquad{\rm if}\ (0)=\chi(0)\ {\rm or}\ \phi(0)\cr
 a&=\gamma_s,\qquad{\rm if}\ (0)=\omega(0)\cr
 a&=\gamma_\sigma,\qquad{\rm if}\ (0)=\varepsilon(0)\cr
\elb

   We are now ready to give the non-relativistic $G^{J}_{\ell s}$'s in terms of 
the relativistic $g^{J}_{\ell s}$'s.  From \eqn{a2a}, \eqn{a10} and \eqn{a14b}, we find
\bln{a12d}
G^{J}_{\ell S}&=\left(\frac{2\ell+1}{2J+1}\right)^{-1/2}
	\!\!\!(\ell0\,S\delta|J\delta)^{-1}\cr
   &\hskip18mm\times
      \sum_i\,^{(i)\!}g^{J}_{\ell S}\;^{(i)\!}A^J_{\ell S}(s,\sigma;\delta)
	\Big|_{\tst(w=w_{\rm th}\;\mbox{and}\;\gamma_s=\gamma_\sigma=1)}
\eln
where $w_{\rm th}$ is the $w$ at its threshold.  
We emphasize here that our decay amplitudes, constructed with the prescription following
\eqn{a11}, correspond to the process $|Jm\ket\to|S\delta\ket+|\ell\,0\ket$ and hence it must be
proportional to the Clebsch-Gordan coefficient $(\ell\,0\;S\delta|J\,\delta)$.  What is remarkable is that
the proportionality constant is independent of $\delta$ in the nonrelativistic limit, 
i.e. $\gamma_s=\gamma_\sigma=1$ and $w=w_{\rm th}$.  This result can be traced to the fact that 
the wave functions $\phi^*(J\delta)$ and $\chi(\ell\,0)\,r^\ell$ are at rest in the $J$RF, so that
they do not have {\em time} components.  So, even though the wave functions $\phi(s,\sigma;S\delta)$
with a total intrinsic spin $S$ have nonzero {\em time} components,  they do not contribute to the
covariant decay amplitudes \eqn{a11} and \eqn{a11a}.  
The resulting amplitudes are proportional to $(w/w_{\rm th})^{n_0}$ with
$n_0\,(=0,\,1)$ defined according to the prescription following \eqn{a11} and are functions of
$\gamma_s$ and $\gamma_\sigma$ only.  As a consequence, our invariant amplitudes lead to the nonrelativistic
helicity-coupling amplitude $F^J_{\lambda\nu}$ of \eqn{a1} through the formula \eqn{a12d}.

We note that, if the Clebsch-Gordan coefficient
in the denominator in \eqn{a12d} is zero, the amplitude $A^J_{\ell S}(s,\sigma;\delta)$ is zero, and
hence $g^{J}_{\ell s}$ and $G^{J}_{\ell s}$ can be set to zero as well.
From \eqn{a11}, we see that $G^{J}_{\ell s}$'s
are proportional to a factor $(i\,w_{\rm th})$ if $s+\sigma+\ell-J=$ odd, but it
is absent if $s+\sigma+\ell-J=$ even.  If we had used the unitless prescription as given \eqn{a11a},
then the factor $(i\,w_{\rm th})$ would have been replaced by $(w/w_{\rm th})$.  
See the section on Conclusions and Discussions for further comments on \eqn{a12d}.
   
   Finally, we wish to point out that there exists one important class of decay modes for which 
there is just one intermediate state with nonzero spin.  
Specifically, let $\sigma=\nu=0$.  Then we have
$S=s$, $\delta=\lambda$ and $\psi(s,\sigma;S\delta)=\omega(s\delta)$, so that
\bln{a12e}
   F^J_{\;\delta}&=\sum_\ell\,g^J_{\ell s}\,A^J_{\ell s}(s,\sigma;\delta)\,r^\ell\cr
    A^J_{\ell s}(s,\sigma;\delta)&=[p^{n_0},\;\omega(s\delta),\;\chi(\ell0),\;\phi^*(J\delta)]_w,
		\quad n_0 =0,\;1\cr
\eln
Here the invariant amplitude refers to the transition 
$|J\delta\ket\to|s\delta\ket+|\ell0\ket$.  The helicity-coupling
amplitudes \eqn{a1} reduce to
\bln{a12f}
F^J_{\;\delta}=\sum_\ell\left(\frac{2\ell+1}{2J+1}\right)^{1/2}\!\!
(\ell0\,s\delta|J\delta)\;G^{J}_{\ell s}\,r^\ell
\eln
The formula \eqn{a12d} for this case reduces to
\bln{a12g}
G^{J}_{\ell s}&=\left(\frac{2\ell+1}{2J+1}\right)^{-1/2}
	\!\!\!(\ell0\,s\delta|J\delta)^{-1}\;
      g^{J}_{\ell s}\;A^J_{\ell s}(s,\sigma;\delta)
	\Big|_{\tst(w=w_{\rm th}\;\mbox{and}\;\gamma_s=1)}
\eln
It can be shown that there exists a single $A^J_{\ell s}(s,\sigma;\delta)$ for any given set
$\{J,\ell,s\}$, i.e. the summation on $i$ in \eqn{a12d} does not apply here.
Once again, the right-hand side of the formula above does not depend on $\delta$,
because its dependence on the amplitude $A^J_{\ell s}(s,\sigma;\delta)$ is cancelled 
by the Clebsch-Gordan coefficient in the denominator.
\sectn{Illustrative Examples I}{sc7}
\indnt
Here we give a set of selected examples to illustrate the principles involved
and to demonstrate the need to carefully examine each, so as to cast the
invariant amplitudes as general as feasible.  
We confine ourselves to the
cases in which both of the decay products have spins given by  $s=1$ or 2 and $\sigma=0$.
Although the decay modes in which one of the two decay products
is spinless have already been treated adequately in a previous paper\cite{SCh0},
we start with three such examples in order to contrast with those treated in Appendix C.
A nontrivial case with $J=3$ is treated as a fourth example.

   We first give in Tables IIa and IIb the decay amplitudes with $s=1$ or 2 and $\sigma=0$
and then give $F^J$'s explicitly with four examples.
\def\arraystretch{1.5}
\begin{center}\hbox{\hbox{\begin{tabular}
{|@{\hspace{18pt}}r@{\hspace{18pt}}|@{\hspace{18pt}}
cc@{\hspace{8pt}}c@{\hspace{8pt}}c@{\hspace{8pt}}c@{\hspace{8pt}}cc@{\hspace{24pt}}|}
\mlt{8}{c}{Table IIa: \vtop{
	\hbox{\undl{Decay Amplitudes}}
	\hbox{for $s+\ell-J=$ even}
	\vskip3pt\hbox{($s=1$ or 2 and $\sigma=0$)}
	\vskip3pt\hbox{$R$ stands for the rank.}
\vskip6pt}}\\
\mlt{2}{c}{}& $R_s(n_1)$ && $R_\ell$ && $R_J$ &\mlt{1}{c}{}\\   
\hline
1           &          & 1(0)        &$\cdot$     
					& 1     &       & 0    &\\
2           &  & 1(1)        &     
					& 0     &       & 1    &\\
3           &          & 1(0)        &$\cdot$     
					& 2     &$\cdot$& 1    &\\
4           &          & 1(1)        &     
					& 1     &$\cdot$& 2    &\\
5           &          & 1(0)        &$\cdot$     
					& 3     &$:$    & 2    &\\
6           &          & 1(1)        &     
					& 2     &$:$    & 3    &\\
7           &          & 1(0)        &$\cdot$     
					& 4     &$:\!\cdot$&3    &\\
8           &          & 1(1)        &     
					& 3     &$\cdot\!:$&4    &\\
9           &          & 1(1)        &$\cdot$     
					& 5     &$::$      &4    &\\
10           &         & 1(1)        &     
					& 4     & $::$    & 5    &\\
11          &          & 1(0)        &$\cdot$     
					& 6     & $::\!\cdot$       & 5    &\\
\hline
1           &          & 2(0)        &$:$     
					& 2     &       & 0    &\\
2           &  & 2(1)        &$\cdot$    
					& 1     &       & 1    &\\
\hline
\end{tabular}}
\hskip24pt\hbox{\begin{tabular}
{|@{\hspace{18pt}}r@{\hspace{18pt}}|@{\hspace{18pt}}
cc@{\hspace{8pt}}c@{\hspace{8pt}}c@{\hspace{8pt}}c@{\hspace{8pt}}cc@{\hspace{24pt}}|}
\mlt{8}{c}{Table IIb: \vtop{
	\hbox{\undl{Decay Amplitudes}}
	\hbox{for $s+\ell-J=$ even}
	\vskip3pt\hbox{($s=1$ or 2 and $\sigma=0$)}
	\vskip3pt\hbox{$R$ stands for the rank.}
\vskip6pt}}\\
\mlt{2}{c}{}& $R_s(n_1)$ && $R_\ell$ && $R_J$ &\mlt{1}{c}{}\\   
\hline
3           &          & 2(0)        &$:$     
					& 3     &$\cdot$       & 1    &\\
4           &      & 2(2)        &     
					& 0     &       & 2    &\\
5           &      & 2(1)        &$\cdot$     
					& 2     &$\cdot$       & 2    &\\
6           &          & 2(0)        &$:$     
					& 4     &$:$    & 2    &\\
7           &      & 2(2)        &     
					& 1     &$\cdot$& 3    &\\
8           &  & 2(1)       &$\cdot$     
					& 3     &$:$    & 3    &\\
9           &   & 2(0)        &$:$     
					& 5     &$:\!\cdot$    & 3    &\\
10           &       & 2(2)        &     
					& 2     &$:$    & 4    &\\
11           &  & 2(1)        &$\cdot$    
					& 4     &$\cdot\!:$&4    &\\
12           &   & 2(0)        &$:$    
					& 6     &$::$&4    &\\
13           &       & 2(2)        &     
					& 3     &$\cdot\!:$&5    &\\
14          &  & 2(1)        &$\cdot$     
					& 5     &$::$      &5    &\\
15          &   & 2(0)        &$:$     
					& 7     &$::\!\cdot$      &5    &\\
\hline
\end{tabular}}}\end{center}
\begin{center}\begin{tabular}
{|@{\hspace{18pt}}r@{\hspace{18pt}}|@{\hspace{18pt}}
cc@{\hspace{8pt}}c@{\hspace{8pt}}c@{\hspace{8pt}}c@{\hspace{8pt}}cc@{\hspace{24pt}}|}
\mlt{8}{c}{Table IIc: \vtop{
	\hbox{\undl{Decay Amplitudes}}
	\hbox{for $s+\ell-J=$ odd}
	\vskip3pt\hbox{($s=1$ or 2 and $\sigma=0$)}
	\vskip3pt\hbox{$R$ stands for the rank.}
\vskip6pt}}\\
\mlt{2}{c}{}& $R_s(n_1)$ && $R_\ell$ && $R_J$ &\mlt{1}{c}{}\\   
\hline
1           &          & 1$(1')$        &     
					& $1'$     &         & $1'$    &\\
2           &          & 1$(1')$        &     
					& 2$'$     &$\cdot$  & 2$'$    &\\
3           &          & 1$(1')$        &     
					& 3$'$     &$:$      & 3$'$    &\\
4           &          & 1$(1')$        &     
					& 4$'$     &$:\!\cdot$&4$'$    &\\
5           &          & 1$(1')$        &     
					& 5$'$     &$::$      &5$'$    &\\
\hline
1           &          & 2$(1')$        &$\cdot$     
					& 2$'$     &         & 1$'$    &\\
2           &         & 2$(2')$        &     
					& 1$'$     &  & 2$'$    &\\
\hline
\end{tabular}
\hskip24pt\begin{tabular}
{|@{\hspace{18pt}}r@{\hspace{18pt}}|@{\hspace{18pt}}
cc@{\hspace{8pt}}c@{\hspace{8pt}}c@{\hspace{8pt}}c@{\hspace{8pt}}cc@{\hspace{24pt}}|}
\mlt{8}{c}{Table IId: \vtop{
	\hbox{\undl{Decay Amplitudes}}
	\hbox{for $s+\ell-J=$ odd}
	\vskip3pt\hbox{($s=1$ or 2 and $\sigma=0$)}
	\vskip3pt\hbox{$R$ stands for the rank.}
\vskip6pt}}\\
\mlt{2}{c}{}& $R_s(n_1)$ && $R_\ell$ && $R_J$ &\mlt{1}{c}{}\\   
\hline
3           &          & 2$(1')$        &$\cdot$     
					& 3$'$     &$\cdot$  & 2$'$    &\\
4           &          & 2$(2')$        &     
					& 2$'$     &$\cdot$      & 3$'$    &\\
5           &          & 2$(1')$        &$\cdot$     
					& 4$'$     &$:$      & 3$'$    &\\
6           &          & 2$(2')$        &     
					& 3$'$     &$:$&4$'$    &\\
7           &          & 2$(1')$        &$\cdot$     
					& 5$'$     &$:\!\cdot$&4$'$    &\\
8           &          & 2$(2')$        &     
					& 4$'$     &$:\!\cdot$      &5$'$    &\\
9           &          & 2$(1')$        &$\cdot$     
					& 6$'$     &$::$      &5$'$    &\\
\hline
\end{tabular}
\end{center}

\subsection{\undl{$1^-\to 1^-+0^-$}}
\indnt
Here we must have $\ell=$ 1 from parity conservation, 
and the decay amplitude \eqn{a11} in the $J\,$RF is
\bln{d3c}
   A^{(1)}_{11}(s,\sigma;\delta)=[\,p,\,\omega(1\,\delta),\,\chi(1\,0),\,\phi^*(1\,\delta)\,]_w
   	&=\epsilon^{\alpha\beta\gamma\delta}\,p_\alpha\,
	\omega_\beta(1\,\delta)\,\chi_\gamma(1\,0)\,\phi^*_\delta(1\,\delta)\cr
	&=w\,\epsilon_{ijk}\,\omega_i(1\,\delta)\,\chi_j(1\,0)\,\phi^*_k(1\,\delta)\cr
	&=\pm(i\,w),\quad\delta=\pm1\cr
	&=0,\quad\delta=0
\eln
The amplitude given here corresponds to row 1, Table IIc.
Note that \eqn{d3c} is in fact the {\em only acceptable}
decay amplitude which {\em satisfies} \eqn{a11x0}.  The helicity-coupling amplitudes are,
from \eqn{a10},
\bln{b0b}
 F^{(1)}_{\;\pm}&=\pm(i\,w)\,g^{(1)}_{11}\,r,\qquad
 F^{(1)}_{\;0}=0
\eln
where $g^{(1)}_{11}$, with $\{\ell\,S\}=\{1\,1\}$, is an arbitrary phenomenological constant.
In the non-relativistic limit, we have $ \sqrt{2}F^{(1)}_{\;\pm}=\mp G^{(1)}_{11}\,r$ 
and $F^{(1)}_{\;0}=0$.
So we conclude $G^{(1)}_{11}=-\sqrt{2}(i\,w)\,g^{(1)}_{11}$.
\subsection{\undl{$2^-\to 1^-+0^-$}}
\indnt
There are two orbital angular momenta $\ell=1$ and $\ell=3$.
The covariant amplitudes are given in rows 4 and 5, Table IIa,
for pure $P$- and $F$-waves, respectively.  They are
\bln{Apta}
   A^{(2)}_{11}(s,\sigma;\delta)&=[\,\omega(1\,\delta),\,\chi(1\,0),\,\phi^*(2\,\delta)\,]_w\cr
	&=\omega^\alpha(1\,\delta)\,\chi^\beta(1\,0)\,\phi^*_{\alpha\beta}(2\,\delta)\cr
   A^{(2)}_{31}(s,\sigma;\delta)&=[\,\omega(1\,\delta),\,\chi(3\,0),\,\phi^*(2\,\delta)\,]_w\cr
 &=\omega_\alpha(1\,\delta)\,\chi^{\alpha\beta\gamma}(3\,0)\,\phi^*_{\beta\gamma}(2\,\delta)\cr
\eln
so that, using the Lorentz metric $\tilde g(w)$ in the $J\,$RF,
\bln{Aptb}
A^{(2)}_{11}(s,\sigma;1)&=\frac{1}{\sqrt{2}},\quad 
	A^{(2)}_{11}(s,\sigma;0)=\sqrt{\frac{2}{3}}\,\gamma_s\cr
A^{(2)}_{31}(s,\sigma;1)&=-\frac{\sqrt{2}}{5},\quad 
	A^{(2)}_{31}(s,\sigma;0)=\frac{\sqrt{6}}{5}\,\gamma_s\cr
\eln
From this one obtains
\blb{Aptc}
 F^{(2)}_+&=\frac{1}{\sqrt{2}} \left(g^{(2)}_{11}-\frac{2}{5}\,g^{(2)}_{31}\,r^2\right)r\cr
 F^{(2)}_0&=\sqrt\frac{2}{3}\gamma_s
   \left(g^{(2)}_{11}+\frac{3}{5}\,g^{(2)}_{31}\,r^2\right)r
\elb
The helicity-coupling amplitudes in the non-relativistic limit are, from \eqn{a12f},
\blb{Ap1}
   \sqrt{2}F^{(2)}_+&=\sqrt\frac{3}{5}G^{(2)}_{11}\,r+\sqrt\frac{2}{5}G^{(2)}_{31}\,r^3\cr
   F^{(2)}_0&=\sqrt\frac{2}{5}G^{(2)}_{11}\,r-\sqrt\frac{3}{5}G^{(2)}_{31}\,r^3
\elb
We obtain, from \eqn{a12g},
\bln{Gg3}
 G^{(2)}_{11}=\sqrt\frac{5}{3}\,g^{(2)}_{11},\quad
 G^{(2)}_{31}=-\sqrt\frac{2}{5}\,g^{(2)}_{31}
\eln
\subsection{\undl{$1^+\to 2^++0^-$}}
\indnt
   This decay can proceed via two allowed values of $\ell$, i.e. $\ell=1$ or 3 
(see rows 2 and 3, table IIb).
\bln{b1a}
   A^{(1)}_{12}(s,\sigma;\delta)&=[\,\omega(2\,\delta),\,\chi(1\,0),\,\phi^*(1\,\delta)\,]_w\cr
	&=\omega^{\alpha\beta}(2\,\delta)\,\chi_\beta(1\,0)\,\phi^*_\alpha(1\,\delta)\cr
	&=\frac{1}{\sqrt{2}}\gamma_s,\quad \delta=+1\cr
	&=\sqrt{\frac{2}{3}}\gamma_s^2,\quad \delta=0\cr
\eln
and
\bln{b1b}
   A^{(1)}_{32}(s,\sigma;\delta)&=[\,\omega(2\,\delta),\,\chi(3\,0),\,\phi^*(1\,\delta)\,]_w\cr
	&=\omega_{\alpha\beta}(2\,\delta)\,\chi^{\alpha\beta\gamma}(3\,0)
		\,\phi^*_\gamma(1\,\delta)\cr
	&=-\frac{\sqrt{2}}{5}\gamma_s,\quad \delta=+1\cr
	&=\frac{1}{5}\sqrt{\frac{2}{3}}(1+2\gamma_s^2),\quad \delta=0\cr
\eln
The helicity-coupling amplitudes are, from \eqn{a10} with $J=1$,
\blb{b1c}
 F^{(1)}_{\;+}&=
	\frac{1}{\sqrt{2}}\left(g^{(1)}_{12}\,r-\frac{2}{5}\;g^{(1)}_{32}\,r^3\right)\gamma_s\cr
 F^{(1)}_{\;0}&=\sqrt{\frac{2}{3}}\left[g^{(1)}_{12}\,\gamma_s^2\,r
	+\frac{3}{5}\,g^{(1)}_{32}\,\left(\frac{1}{3}+\frac{2}{3}\gamma_s^2\right)\,r^3\right]\cr
\elb
where $g^{(1)}_{12}$ and $g^{(1)}_{32}$ are arbitrary phenomenological constants.

    In the non-relativistic limit, we obtain
\blb{b1d}
  \sqrt{2}F^{(1)}_{\;+}&=
	-\sqrt{\frac{3}{5}}\,G^{(1)}_{12}\,r-\sqrt{\frac{2}{5}}\,G^{(1)}_{32}\,r^3\cr
  F^{(1)}_{\;0}&=-\sqrt{\frac{2}{5}}\,G^{(1)}_{12}\,r+\sqrt{\frac{3}{5}}\,G^{(1)}_{32}\,r^3
\elb
Here we see for the first time the rules \eql{a1x0}{a} and \eql{a1x0}{b} are obeyed by the
formula above.  Note also that \eqn{a1x1} is satisfied as well.  If we set $\gamma_s=1$ in
\eqn{b1c}, we find that 
\bln{b1e}
G^{(1)}_{12}=-\sqrt{\frac{5}{3}}\,g^{(1)}_{12}\quad \mbox{and}\quad 
	G^{(1)}_{32}=\sqrt{\frac{2}{5}}\,g^{(1)}_{32}
\eln
which can be obtained directly by applying \eqn{a12g}.
\subsection{\undl{$3^+\to 2^++0^-$}}
\indnt
   We give, as a final example of this section, 
the helicity-coupling amplitudes for $J=3$, $s=2$ and $\sigma=0$.
The requisite amplitudes, $A^{(3)}_{\ell\,S}(s,\sigma;\delta)$, require three
orbital angular momenta $\ell=1$, 3 and 5, as given 
by rows 7, 8, and 9 in Table IIb.  The results are
\bln{j3x}
A^{(3)}_{12}(s,\sigma;2)&=\frac{1}{\sqrt{3}},\; 
A^{(3)}_{12}(s,\sigma;1) =2\sqrt{\frac{2}{15}}\;\gamma_s,\;
A^{(3)}_{12}(s,\sigma;0) =\frac{1}{\sqrt{15}}\;(2\gamma^2_s+1)\cr
A^{(3)}_{32}(s,\sigma;2)&=-\frac{2}{5\sqrt{3}},\;
A^{(3)}_{32}(s,\sigma;1) =\frac{1}{5}\sqrt{\frac{2}{15}}\;\gamma_s,\;
A^{(3)}_{32}(s,\sigma;0) =\frac{2}{15}\sqrt{\frac{3}{5}}\;(3\gamma^2_s-1)\cr
A^{(3)}_{52}(s,\sigma;2)&=\frac{2}{21\sqrt{3}},\; 
A^{(3)}_{52}(s,\sigma;1) =-\frac{10}{21}\sqrt{\frac{2}{15}}\;\gamma_s,\;
A^{(3)}_{52}(s,\sigma;0) =\frac{20}{63\sqrt{15}}\;(2\gamma^2_s+1)\cr
\eln
which lead to
\blb{j3o}
   F^{(3)}_2&=\frac{1}{\sqrt{3}} \ \left(g^{(3)}_{12}-\frac{2}{5}g^{(3)}_{32}\,r^2
                                   +\frac{2}{21}g^{(3)}_{52}\,r^4\right)r\cr
%
%
   F^{(3)}_1&=\sqrt{\frac{\phantom{!}2}{15}}
         \left(2g^{(3)}_{12}+\frac{1}{5}g^{(3)}_{32}\,r^2
               -\frac{10}{21}g^{(3)}_{52}\,r^4\right)\;\gamma_s\;r\cr
   F^{(3)}_0&=
    \sqrt\frac{3}{5}\ \Biggl\{g^{(3)}_{12}\,\left(\frac{2}{3}\gamma_s^2+\frac{1}{3}\right)
   +\frac{4}{15}g^{(3)}_{32}\,\left(\frac{3}{2}\gamma_s^2-\frac{1}{2}\right)r^2
   +\frac{20}{63}g^{(3)}_{52}\left(\frac{2}{3}\gamma_s^2+\frac{1}{3}\right)r^4\Biggr\}r\cr
\elb
where $g^{(3)}_{12}$, $g^{(3)}_{32}$ and $g^{(3)}_{52}$ are the constants in the problem, 
corresponding to $\ell=1$, 3 and 5.

   In the non-relativistic limit, we find that
\blb{j3p}
   \sqrt{2}F^{(3)}_2&=\sqrt{\frac{2}{7}}\,G^{(3)}_{12}\,r+\sqrt{\frac{2}{3}}\,G^{(3)}_{32}\,r^3
		+\sqrt{\frac{1}{21}}\,G^{(3)}_{52}\,r^5\cr
   \sqrt{2}F^{(3)}_1&=\frac{4}{\sqrt{35}}\,G^{(3)}_{12}\,r-\sqrt{\frac{1}{15}}\,G^{(3)}_{32}\,r^3
		-\sqrt{\frac{10}{21}}\,G^{(3)}_{52}\,r^5\cr
   F^{(3)}_0&=\frac{3}{\sqrt{35}}\,G^{(3)}_{12}\,r-\frac{2}{\sqrt{15}}\,G^{(3)}_{32}\,r^3
		+\sqrt{\frac{10}{21}}\,G^{(3)}_{52}\,r^5\cr
\elb
Setting $\gamma_s=1$ in \eqn{j3o}, we see that 
\bln{j3q}
G^{(3)}_{12}=\sqrt{\frac{7}{3}}\;g^{(3)}_{12},\quad
	G^{(3)}_{32}=-\frac{2}{5}\;g^{(3)}_{32},\quad\mbox{and}\quad 
		G^{(3)}_{52}=\frac{2}{3}\sqrt{\frac{2}{7}}\;g^{(3)}_{52}
\eln
This can be obtained directly from \eqn{a12g}. 
\sectn{Illustrative Examples II}{sc8}
\indnt
Here we confine ourselves to the
cases in which both of the decay products have spins given by  $s=1$ and $\sigma=1$.
The corresponding decay amplitudes are worked out in Tables IIIa and IIIb.
To further illustrate the techniques involved, $F^J$'s are explicitly worked out
with a few examples.
\def\arraystretch{1.5}
\begin{center}\hbox{\hbox{\begin{tabular}
{|@{\hspace{4pt}}r@{\hspace{6pt}}|@{\hspace{6pt}}l@{\hspace{6pt}}|@{\hspace{-6pt}}
cc@{\hspace{4pt}}c@{\hspace{4pt}}c@{\hspace{4pt}}c@{\hspace{4pt}}cc@{\hspace{-6pt}}|}
\mlt{8}{c}{Table IIIa: \vtop{
	\hbox{\undl{Decay Amplitudes}}
	\hbox{for $\ell-J=$ even}	
	\vskip3pt\hbox{($s=1$ and $\sigma=1$)}
	\vskip3pt\hbox{$R$ stands for the rank.}
\vskip6pt}}\\
\hline
&$S$&& $R_S(n_1,n_2)$ && $R_\ell$ && $R_J$ &\\   
\hline
1&0           &          &$\onedot{2}$$(0,0)$        &       
					& 0     &       & 0    &\\
2&2           &          &           2$(0,0)$        &   $:$    
					& 2     &       & 0    &\\
\hline
3&$0,\;1,\;2$ &          &$\onedot{2}$$(0,0)$        &       
					& 1     &$\cdot$& 1    &\\
4&$0,\;1,\;2$&    &           2$(1,0)$        &   $\cdot$    
					& 1     &       & 1    &\\
5&$0,\;1,\;2$&   &           2$(0,1)$        &   $\cdot$    
					& 1     &       & 1    &\\
6&2          &           &           2$(0,0)$        &   $:$    
					& 3     &$\cdot$& 1    &\\
\hline
7&2          &       &           2$(1,1)$        &
					& 0     &       & 2    &\\
8&$0,\;1,\;2$ &          &$\onedot{2}$$(0,0)$        &       
					& 2     &$:$    & 2    &\\
9&$0,\;1,\;2$&   &           2$(1,0)$        &   $\cdot$    
					& 2     &$\cdot$& 2    &\\
10&$0,\;1,\;2$&   &           2$(0,1)$        &   $\cdot$    
					& 2     &$\cdot$& 2    &\\
11&2          &           &           2$(0,0)$        &   $:$    
					& 4     &$:$    & 2    &\\
\hline
12&$2$        &    &           2$(1,1)$        &
					& 1     &$\cdot$& 3    &\\
13&$0,\;1,\;2$&           &  $\onedot{2}(0,0)$        & 
					& 3     &$:\!\cdot$& 3    &\\
\hline
\end{tabular}}
\hskip36pt\hbox{\begin{tabular}
{|@{\hspace{4pt}}r@{\hspace{6pt}}|@{\hspace{6pt}}l@{\hspace{6pt}}|@{\hspace{-6pt}}
cc@{\hspace{4pt}}c@{\hspace{4pt}}c@{\hspace{4pt}}c@{\hspace{4pt}}cc@{\hspace{-6pt}}|}
\mlt{8}{c}{Table IIIa: \vtop{
	\hbox{\undl{Decay Amplitudes}}
	\hbox{for $\ell-J=$ even}	
	\vskip3pt\hbox{($s=1$ and $\sigma=1$)}
	\vskip3pt\hbox{$R$ stands for the rank.}
\vskip6pt}}\\
\hline
&$S$&& $R_S(n_1,n_2)$ && $R_\ell$ && $R_J$ &\\   
\hline
14&$0,\;1,\;2$ &  &           2$(1,0)$        & $\cdot$      
					& 3     &$:$    & 3    &\\
15&$0,\;1,\;2$ &  &           2$(0,1)$        & $\cdot$      
					& 3     &$:$    & 3    &\\
16&2          &           &           2$(0,0)$        &   $:$    
					& 5     &$:\!\cdot$    & 3    &\\
\hline
17&2          &$:$        &           2$(1,1)$         &
					& 2   &$:$       & 4    &\\
18&$0,\;1,\;2$ &          &$\onedot{2}$$(0,0)$         &       
					& 4   &$::$      & 4    &\\
19&$0,\;1,\;2$&    &           2$(1,0)$         &   $\cdot$    
					& 4   &$:\!\cdot$& 4    &\\
20&$0,\;1,\;2$&    &           2$(0,1)$         &   $\cdot$    
					& 4   &$:\!\cdot$& 4    &\\
21& 2         &           &           2$(0,0)$         &   $:$    
					& 6   &$::$      & 4    &\\
\hline
22&$2$        &      &           2$(1,1)$         &
					& 3  &$:\!\cdot$ & 5    &\\
23&$0,\;1,\;2$&           &  $\onedot{2}(0,0)$         & 
					& 5  &$::\!\cdot$& 5    &\\
24&$0,\;1,\;2$ &  &           2$(1,0)$         & $\cdot$      
					& 5  &$::$       & 5    &\\
25&$0,\;1,\;2$ &  &           2$(0,1)$         & $\cdot$      
					& 5  &$::$       & 5    &\\
26&2          &           &           2$(0,0)$         &   $:$    
					& 7  &$::\!\cdot$& 5    &\\
\hline
\end{tabular}}}\end{center}
\def\arraystretch{1.5}
\begin{center}\hbox{\hbox{\begin{tabular}
{|@{\hspace{4pt}}r@{\hspace{6pt}}|@{\hspace{6pt}}l@{\hspace{6pt}}|@{\hspace{-6pt}}
cc@{\hspace{4pt}}c@{\hspace{4pt}}c@{\hspace{4pt}}c@{\hspace{4pt}}cc@{\hspace{-6pt}}|}
\mlt{9}{c}{Table IIIb: \vtop{
	\hbox{\undl{Decay Amplitudes}}
	\hbox{for $\ell-J=$ odd}	
	\vskip3pt\hbox{($s=1$ and $\sigma=1$)}
	\vskip3pt\hbox{$R$ stands for the rank.}
\vskip6pt}}\\
\hline
&$S$&& $R_S(n_1,n_2)$ && $R_\ell$ && $R_J$ & \\  
\hline
1&   1   &       &   2$(1',1')$   &       & $1'$     &       & 0    &            \\     
\hline
2&   1   &       &   2$(1',1')$   &       & 0     &       & $1'$    &           \\      
3&1, 2   &       &   2$(1',0)$   &$\cdot$& $2'$     &       & $1'$    &           \\
4&1, 2       &       &   2$(0,1')$   &$\cdot$& $2'$     &       & $1'$    &   \\
5&1, 2       &       &   2$(1',1')$   &       & $2'$     &$\cdot$& 1    &    \\
\hline
6&1, 2&          &   2$(1',1')$   &       & 1     &$\cdot$& $2'$    &    \\
7&1, 2    &  &   2$(1',1)$   &       & $1'$     &       & $2'$    & \\
8&1, 2    &  &   2$(1,1')$   &       & $1'$     &       & $2'$    & \\
9&1, 2    &          &   2$(1',1')$   &       & $3'$     &$:$    & 2    &  \\
10&1, 2    &          &   2$(1',0)$   &$\cdot$& $3'$     &$\cdot$& $2'$    &  \\
11&1, 2    &          &   2$(0,1')$   &$\cdot$& $3'$     &$\cdot$& $2'$    & \\
\hline
12&1, 2&          &   2$(1',1')$   &       & 2     &$:$     & $3'$    &            \\     
13&1, 2    &  &   2$(1',1)$   &       & $2'$     &$\cdot$ & $3'$    &    \\     
14&1, 2    &   &   2$(1,1')$   &       & $2'$     &$\cdot$ & $3'$    &  \\
15&1, 2    &          &   2$(1',0)$   &$\cdot$& $4'$     &$:$     & $3'$    &           \\
\hline
\end{tabular}}
\def\arraystretch{1.5}
\hskip48pt\hbox{\begin{tabular}
{|@{\hspace{4pt}}r@{\hspace{6pt}}|@{\hspace{6pt}}l@{\hspace{6pt}}|@{\hspace{-6pt}}
cc@{\hspace{4pt}}c@{\hspace{4pt}}c@{\hspace{4pt}}c@{\hspace{4pt}}cc@{\hspace{-6pt}}|}
\mlt{9}{c}{Table IIIb: \vtop{
	\hbox{\undl{Decay Amplitudes}}
	\hbox{for $\ell-J=$ odd}	
	\vskip3pt\hbox{($s=1$ and $\sigma=1$)}
	\vskip3pt\hbox{$R$ stands for the rank.}
\vskip6pt}}\\
\hline
&$S$&& $R_S(n_1,n_2)$ && $R_\ell$ && $R_J$ &  \\ 
\hline
16&1, 2    &          &   2$(0,1')$   &$\cdot$& $4'$     &$:$     & $3'$    &           \\
17&1, 2    &          &   2$(1',1')$   &       & $4'$     &$:\!\cdot$& 3    &        \\
\hline
18&1, 2&          &   2$(1',1')$   &       & 3     &$:\!\cdot$& $4'$    &            \\     
19&1, 2    &   &   2$(1',1)$   &       & $3'$     &$:$       & $4'$    &    \\     
20&1, 2    &  &   2$(1,1')$   &       & $3'$     &$:$       & $4'$    &  \\
21&1, 2    &          &   2$(1',1')$   &       & $5'$     &$::$      & 4    &    \\
22&1, 2    &          &   2$(1',0)$   &$\cdot$& $5'$     &$:\!\cdot$& $4'$    &           \\
23&1, 2    &          &   2$(0,1')$   &$\cdot$& $5'$     &$:\!\cdot$& $4'$    &           \\
\hline
24&1, 2&          &   2$(1',1')$   &       & 4     &$::$      & $5'$    &            \\     
25 &1, 2    &  &   2$(1',1)$   &       & $4'$     &$:\!\cdot$&  $5'$    &     \\     
26&1, 2    &  &   2$(1,1')$   &       & $4'$     &$:\!\cdot$&  $5'$    &    \\
27&1, 2    &          &   2$(1',0)$   &$\cdot$& $6'$     &$::$      &  $5'$    &           \\
28&1, 2    &          &   2$(0,1')$   &$\cdot$& $6'$     &$::$      &  $5'$    &           \\
29&1, 2    &          &   2$(1',1')$  &       & $6'$     &$::\!\cdot$& 5    &        \\
\ftm{29}& &    &   \ftm{2$(1',1')$} & & \ftm{$6'$}     && \ftm{5}   &        \\
\hline
\end{tabular}}}\end{center}
\subsection{\undl{$0^+\to 1^-+1^-$}}
\indnt
The allowed orbital angular momenta are $\ell=0$ and 2 (see rows 1 and 2, Table IIIa).  
Since $J=0$, we must have $S=\ell=0$ or 2 and $\lambda=\nu=0$ or 1.
So we find in the $J\,$RF
\bln{c5}
A^{(0)}_{00}(s,\sigma;0)
&=[\onedot\Psi(0,0;00)]_w
=\tilde g_{\alpha\beta}(w)\,\Psi^{\alpha\beta}(0,0;00)\cr
&=\sum_i\,\Psi_{ii}(0,0;00)
=-\frac{1}{\sqrt{3}} (2+\gamma_s\gamma_\sigma)\cr
A^{(0)}_{22}(s,\sigma;0)&=[\Psi(0,0;20):\chi(20)]_w\cr
&=\sum_{ij}\,\Psi_{ij}(0,0;20)\;\chi_{ji}(20)
=\frac{1}{3}\sqrt\frac{2}{3}\,(1+2\gamma_s\gamma_\sigma)\cr
\eln
where we have used the notation $\onedot\psi$ to indicate internal contraction of
the rank-2 tensor.  The helicity-coupling amplitudes are, from \eqn{a10},
\blb{c6a}
F^{(0)}_{\pm\pm}&=-g^{(0)}_{00}\,\left(\frac{1}{3}\gamma_s\gamma_\sigma+\frac{2}{3}\right)
+\frac{1}{3}\,g^{(0)}_{22}\,\left(\frac{2}{3}\gamma_s\gamma_\sigma+\frac{1}{3}\right)\,r^2\cr
F^{(0)}_{\;00}&=g^{(0)}_{00}\,\left(\frac{1}{3}\gamma_s\gamma_\sigma+\frac{2}{3}\right)
+\frac{2}{3}\,g^{(0)}_{22}\,\left(\frac{2}{3}\gamma_s\gamma_\sigma+\frac{1}{3}\right)\,r^2\cr
\elb
We write the amplitudes
so that the elements dependent on Lorentz factors are normalized to 1.
Under the requirement of Bose symmetrization, the amplitudes above must be symmetric
under interchange of $s$ and $\sigma$.  We see that it is automatically satisfied.

Consider next the case of a very large mass for the parent particle.
In the limit $w\to\infty$, we find
\blb{c92}
   F^{(0)}_{\pm\pm}&\to\frac{1}{3}\,\left(-g^{(0)}_{00}
	+\frac{2}{3}\,g^{(0)}_{22}\,r^2\right)\gamma_s\gamma_\sigma
	\to\frac{2}{9}\,g^{(0)}_{22}\gamma_s\gamma_\sigma\,r^2\cr
   F^{(0)}_{00}&\to\frac{1}{3}\,\left(g^{(0)}_{00}
	+\frac{4}{3}\,g^{(0)}_{22}\,r^2\right)\gamma_s\gamma_\sigma
	\to\frac{4}{9}\,g^{(0)}_{22}\gamma_s\gamma_\sigma\,r^2\cr
\elb
since $r^2\to\infty$ in this limit.

In the non-relativistic limit, the helicity-coupling amplitudes are given by
\blb{c93}
 \sqrt{2}F^J_{\pm\pm}&=\sqrt\frac{2}{3}G^J_{00}
               +\sqrt\frac{1}{3}G^J_{22}\,r^2\cr
 F^J_{00}&=-\sqrt\frac{1}{3}G^J_{00}+\sqrt\frac{2}{3}G^J_{22}\,r^2
\elb
with $J=0$. We set $\gamma_s=\gamma_\sigma=1$ and find
\bln{c94}
   G^J_{00}=-\sqrt{3}\;g^J_{00}\quad\mbox{and}\quad
	G^J_{22}=\sqrt{\frac{2}{3}}\;g^J_{22}
\eln
This is consistent with the formula \eqn{a12d}.
\subsection{\undl{$0^-\to 1^-+1^-$}}
\indnt
   The allowed orbital angular momentum is $\ell=1$ (see row 1, Table IIIb). 
Since $J=0$, we must have $S=\ell=1$.
The decay amplitude is
\bln{c9a}
A^{(0)}_{1 1}(s,\sigma;0)&=[\,p,\;\Psi(1',1';10),\;\chi(1'0)\,]_w\,r=\sqrt{2}\,(i\,w)\cr
\eln
and the corresponding helicity-coupling amplitude is
\bln{c9b}
   F^{(0)}_{++}=(i\,w)\,g^{(0)}_{1 1}\,r,\quad  F^{(0)}_{\;00}=0
\eln
Again, the Bose symmetry is automatic.  In the non-relativistic limit, 
we have $\sqrt{2}F^{(0)}_{++}=-G^{(0)}_{1 1}\,r$ and $F^{(0)}_{\;00}=0$.
So we obtain $G^{(0)}_{1 1}=-\sqrt{2}\;(i\,w)\,g^{(0)}_{1 1}$.
\subsection{\undl{$1^-\to 1^++1^-$}}
\indnt
   The allowed orbital angular momenta are $\ell=0$ and 2 (see rows 2, 3, 4 and 5, Table IIIb).  
Since $J=1$, we must have
$S=1$ or 2.  The invariant amplitudes are, with $\delta=\lambda-\nu$,
\bln{c20}   
A^{(1)}_{0 S}(s,\sigma;\delta)
&=[p,\;\Psi(1',1';S\delta),\;\phi^*(1'\delta)]_w ,\quad S=1\cr
A^{(1)}_{2 S}(s,\sigma;\delta)
&=[p,\;\Psi(1',0;S\delta)\cdot\chi(2'0),\;\phi^*(1'\delta)]_w,\quad S=1,2\cr
B^{(1)}_{2 S}(s,\sigma;\delta)
&=[p,\;\Psi(0,1';S\delta)\cdot\chi(2'0),\;\phi^*(1'\delta)]_w
	,\quad S=1,2\cr
C^{(1)}_{2 S}(s,\sigma;\delta)
&=[p,\;\Psi(1',1';S\delta),\;\chi(2'0)\cdot\phi^*(1\delta)]_w
	,\quad S=1,2\cr
\eln
and so
\bln{c20a}
A^{(1)}_{0 1}(s,\sigma;+)&=\frac{1}{\sqrt{2}} \,(i\,w)(\gamma_s+\gamma_\sigma),
\qquad A^{(1)}_{0 1}(s,\sigma;0)=\sqrt{2}\,(i\,w)
\eln
and
\bln{c20b}
A^{(1)}_{2 1}(s,\sigma;+)&=\frac{1}{3\sqrt{2}}\,(i\,w)\,(2\gamma_\sigma-\gamma_s),\quad
B^{(1)}_{2 1}(s,\sigma;+)= - \frac{1}{3\sqrt{2}}\,(i\,w)\,(2\gamma_s-\gamma_\sigma)\cr
%
A^{(1)}_{2 1}(s,\sigma;0)&=-B^{(1)}_{2 1}(0)=-\frac{\sqrt{2}}{3}\,(i\,w)\cr
C^{(1)}_{2 1}(s,\sigma;+)&=-\frac{1}{3\sqrt{2}}\,(i\,w)\,(\gamma_s+\gamma_\sigma),\quad
C^{(1)}_{2 1}(s,\sigma;0)=\frac{2\sqrt{2}}{3}\,(i\,w)\cr
\eln
We find, in addition,
\bln{c20da}
A^{(1)}_{2 2}(s,\sigma;+)&=\frac{(i\,w)}{3\sqrt{2}}(2\gamma_\sigma+\gamma_s),\qquad
A^{(1)}_{2 2}(s,\sigma;0)=0\cr   
\eln
and
\bln{c20e}
B^{(1)}_{2 2}(+)&=\frac{(i\,w)}{3\sqrt{2}}(2\gamma_s+\gamma_\sigma),\qquad
B^{(1)}_{2 2}(0)=0\cr   
C^{(1)}_{2 2}(+)&=-\frac{(i\,w)}{3\sqrt{2}}(\gamma_\sigma-\gamma_s),\qquad   
C^{(1)}_{2 2}(0)=0\cr   
\eln

   So we define the most general helicity-coupling amplitudes to be
\bln{c8a}
   F^J_{\lambda\nu}=F^J_{\lambda\nu}(A)+F^J_{\lambda\nu}(B)+F^J_{\lambda\nu}(C)
\eln
We assign constants $g^J_{\ell S}$'s, $f^J_{\ell S}$'s and $h^J_{\ell S}$'s 
to the amplitudes $A$, $B$ and $C$, respectively, and obtain
\blb{c210}
   F^{(1)}_{++}  &=(i\,w)\,\bigg[g^{(1)}_{\;01}
	-\frac{1}{3}\,g^{(1)}_{\;21}\,r^2  +  \frac{1}{3}f^{(1)}_{\;21}\,r^2
	+\frac{2}{3}h^{(1)}_{\;21}\,r^2\bigg]\cr 
   F^{(1)}_{\;0+}&=(i\,w)\,\bigg[g^{(1)}_{\;01}
	\left(\frac{1}{2}\gamma_\sigma+\frac{1}{2}\gamma_s\right)
	-\frac{1}{3}h^{(1)}_{\;21}\left(\frac{1}{2}\gamma_\sigma+\frac{1}{2}\gamma_s\right)\,r^2\cr
	&\hskip6mm+\frac{1}{6}\,g^{(1)}_{\;21}\left(2\gamma_\sigma-\gamma_s\right)\,r^2
	- \frac{1}{6}\,f^{(1)}_{\;21}\left(2\gamma_s-\gamma_\sigma\right)\,r^2
	+\frac{1}{6}h^{(1)}_{\;22}(\gamma_\sigma-\gamma_s)\,r^2\cr
&\hskip6mm-\frac{1}{2}\,g^{(1)}_{\;22}\left(\frac{2}{3}\gamma_\sigma+\frac{1}{3}\gamma_s\right)\,r^2
	-\frac{1}{2}\,f^{(1)}_{\;22}
	\left(\frac{2}{3}\gamma_s+\frac{1}{3}\gamma_\sigma\right)\,r^2\bigg]\cr
   F^{(1)}_{\;+0}&=(i\,w)\,\bigg[g^{(1)}_{\;01}
	\left(\frac{1}{2}\gamma_\sigma+\frac{1}{2}\gamma_s\right)
	-\frac{1}{3}h^{(1)}_{\;21}\left(\frac{1}{2}\gamma_\sigma+\frac{1}{2}\gamma_s\right)\,r^2\cr
	&\hskip6mm+\frac{1}{6}\,g^{(1)}_{\;21}\left(2\gamma_\sigma-\gamma_s\right)\,r^2
	-\frac{1}{6}\,f^{(1)}_{\;21}\left(2\gamma_s-\gamma_\sigma\right)\,r^2
	-\frac{1}{6}h^{(1)}_{\;22}(\gamma_\sigma-\gamma_s)\,r^2\cr
&\hskip6mm+\frac{1}{2}\,g^{(1)}_{\;22}\left(\frac{2}{3}\gamma_\sigma+\frac{1}{3}\gamma_s\right)\,r^2
	+\frac{1}{2}\,f^{(1)}_{\;22}
	\left(\frac{2}{3}\gamma_s+\frac{1}{3}\gamma_\sigma\right)\,r^2\bigg]
\elb
We conclude that the three helicity-coupling amplitudes $F$ depend on 
a total of seven parameters, three $g$'s, two $f$'s and two $h$'s.
The terms with
$h^{(1)}_{\;22}$ are \emph{purely} relativistic; they vanish 
in the limit $\gamma_s=\gamma_\sigma=1$.

This example can be used to also evaluate $1^+\to1^-+1^-$ with the two vector
states being identical.  The resulting helicity-coupling amplitudes are,
from \eqn{a11x4},
\blb{c21b}
   F^{(1)}_{++}  &=0\cr 
   F^{(1)}_{\;0+}&=(i\,w)\bigg[\frac{1}{4}\,(g^{(1)}_{\;21}-f^{(1)}_{\;21})
			\left(\gamma_\sigma-\gamma_s\right)
	-\frac{1}{2}\,(g^{(1)}_{\;22}-f^{(1)}_{\;22})\,
		\left(\frac{1}{2}\gamma_\sigma+\frac{1}{2}\gamma_s\right)\bigg]\,r^2\cr
   F^{(1)}_{\;+0}&=(i\,w)\bigg[\frac{1}{4}\,(g^{(1)}_{\;21}-f^{(1)}_{\;21})
			\left(\gamma_\sigma-\gamma_s\right)
	+\frac{1}{2}\,(g^{(1)}_{\;22}-f^{(1)}_{\;22})\,
		\left(\frac{1}{2}\gamma_\sigma+\frac{1}{2}\gamma_s\right)\bigg]\,r^2\cr
\elb
and the amplitudes depend only on two parameters $(g^{(1)}_{\;21}-f^{(1)}_{\;21})$
and $(g^{(1)}_{\;22}-f^{(1)}_{\;22})$.  The first term is {\em purely} relativistic
and so it vanishes in the limit $\gamma_s=\gamma_\sigma=1$.
In addition, one may expect that 
the first term should remain small and insignificant, since $\gamma_s$
should be nearly equal to $\gamma_\sigma$ independent of the size of $w$ compared to
$m$ or $\mu$.
   
   The helicity-coupling amplitudes are, in the non-relativistic limit,
\blb{c21}
   \sqrt{2}F^{(1)}_{++}&=\frac{1}{\sqrt{3}}G^{(1)}_{01}
   -\sqrt\frac{2}{3}\ G^{(1)}_{21}\,r^2\cr
   \sqrt{2}F^{(1)}_{\;0+}&=\frac{1}{\sqrt{3}}G^{(1)}_{01}
   +\frac{1}{\sqrt{6}} G^{(1)}_{21}\,r^2-\frac{1}{\sqrt{2}}  G^{(1)}_{22}\,r^2\cr
   \sqrt{2}F^{(1)}_{+0}&=\frac{1}{\sqrt{3}}G^{(1)}_{01}
   +\frac{1}{\sqrt{6}} G^{(1)}_{21}\,r^2+\frac{1}{\sqrt{2}}  G^{(1)}_{22}\,r^2
\elb
So there are just three parameters in this limit.  Note that the summations indicated
by \eql{a1x0}{a} and \eql{a1x0}{b} hold for the formula above.
By setting $\gamma_s=\gamma_\sigma=1$,
we obtain
\blgb{24pt}{c22}
   G^{(1)}_{01}&=\sqrt{6}\,(i\,w)\,g^{(1)}_{01}, &
   G^{(1)}_{21}&=\frac{1}{\sqrt{3}}\,(i\,w)\,\left(g^{(1)}_{21}-f^{(1)}_{21}-2h^{(1)}_{21}\right)\cr
   G^{(1)}_{22}&=(i\,w)\,\left(g^{(1)}_{22}+f^{(1)}_{22}\right)\cr
\elgb
Or, alternatively, we could have used \eqn{a12d} to obtain the same result.
\subsection{\undl{$1^+\to 1^++1^-$}}
\indnt
The allowed orbital angular momenta are $\ell=1$ and 3 (see rows 3, 4, 5 and 6, Table IIIa).
Since $J=1$, we can have
$S=0$, 1 or 2.  The invariant amplitudes are, with $\delta=\lambda-\nu$,
\blb{d0}   
A^{(1)}_{1 S}(s,\sigma;\delta)
&=[\Psi(1,0;S\delta)\cdot\chi(10)\hskip6pt\phi^*(1\delta)]_w,\quad
S=0,\;1,\;2\cr
B^{(1)}_{1 S}(s,\sigma;\delta)
&=[\Psi(0,1;S\delta)\cdot\chi(10)\hskip6pt\phi^*(1\delta)]_w,\quad
S=0,\;1,\;2\cr
C^{(1)}_{1 S}(s,\sigma;\delta)
&=[\,\onedot\Psi(0,0;S\delta)\hskip6pt\chi(10)\cdot\phi^*(1\delta)\,]_w,\quad
S=0,\;1,\;2\cr
A^{(1)}_{3 S}(s,\sigma;\delta)
&=[\Psi(0,0;S\delta)\,:\,\chi(30)\cdot\phi^*(1\delta)]_w,\quad
S=2\cr
\elb
We find
\bln{d2}
   A^{(1)}_{1 0}(s,\sigma;0)&=-\frac{1}{\sqrt{3}} \gamma_s\gamma_\sigma,\quad
   	B^{(1)}_{1 0}(0)=-\frac{1}{\sqrt{3}} \gamma_s\gamma_\sigma
   		\quad C^{(1)}_{1 0}(0)=-\frac{1}{\sqrt{3}} (2+\gamma_s\gamma_\sigma)\cr
   A^{(1)}_{1 1}(s,\sigma;0)&=0,\quad B^{(1)}_{1 1}(0)=0,\quad C^{(1)}_{1 1}(0)=0\cr
   A^{(1)}_{1 1}(s,\sigma;+)&=-\frac{1}{\sqrt{2}} \gamma_s,\quad
   	B^{(1)}_{1 1}(s,\sigma;+)=+\frac{1}{\sqrt{2}} \gamma_\sigma
		\quad C^{(1)}_{1 1}(+)=0\cr
   A^{(1)}_{1 2}(s,\sigma;0)&=\sqrt\frac{2}{3}\;\gamma_s\gamma_\sigma,\quad
   	B^{(1)}_{1 2}(s,\sigma;0)=\sqrt\frac{2}{3}\;\gamma_s\gamma_\sigma
		\quad C^{(1)}_{1 2}(s,\sigma;0)=-\sqrt\frac{2}{3}\,(1-\gamma_s\gamma_\sigma)\cr
   A^{(1)}_{1 2}(s,\sigma;+)&=\frac{1}{\sqrt{2}} \gamma_s,\quad
   	B^{(1)}_{1 2}(s,\sigma;+)=\frac{1}{\sqrt{2}} \gamma_\sigma\,r,\quad C^{(1)}_{1 2}(+)=0\cr
\eln
and
\bln{d3}
   A^{(1)}_{3 2}(s,\sigma;0)&=\frac{1}{5}\sqrt\frac{2}{3}(1+2\gamma_s\gamma_\sigma),\quad
   	A^{(1)}_{3 2}(s,\sigma;+)=-\frac{1}{5\sqrt{2}}(\gamma_s+\gamma_\sigma)\cr
\eln
We assign constants $g^J_{\ell S}$'s, $f^J_{\ell S}$'s and $h^J_{\ell S}$'s 
to the amplitudes $A$, $B$ and $C$, respectively, and obtain
\blb{d4}
   F^{(1)}_{++}&=-\frac{1}{3}\left[(g^{(1)}_{10}+f^{(1)}_{10})\,\gamma_s\gamma_\sigma
	+3h^{(1)}_{10}\,\left(\frac{2}{3}+\frac{1}{3}\gamma_s\gamma_\sigma\right)\right]\,r\cr
&\hskip12mm+\frac{1}{3}\left[(g^{(1)}_{12}+f^{(1)}_{12})\,\gamma_s\gamma_\sigma
	+h^{(1)}_{12}\,(1-\gamma_s\gamma_\sigma)\right]\,r
	+\frac{1}{5}g^{(1)}_{32}\,\left(\frac{1}{3}+\frac{2}{3}\gamma_s\gamma_\sigma\right)\,r^3\cr
   F^{(1)}_{\;0+}&=\frac{1}{2}(g^{(1)}_{11}\,\gamma_s-f^{(1)}_{11}\,\gamma_\sigma)\,r
	+\frac{1}{2}(g^{(1)}_{12}\,\gamma_s+f^{(1)}_{12}\,\gamma_\sigma)\,r
	-\frac{1}{5}g^{(1)}_{32}\,\left(\frac{1}{2}\gamma_s+\frac{1}{2}\gamma_\sigma\right)\,r^3\cr
   F^{(1)}_{\,+0}&=-\frac{1}{2}(g^{(1)}_{11}\,\gamma_s-f^{(1)}_{11}\,\gamma_\sigma)\,r
	+\frac{1}{2}(g^{(1)}_{12}\,\gamma_s+f^{(1)}_{12}\,\gamma_\sigma)\,r
	-\frac{1}{5}g^{(1)}_{32}\,\left(\frac{1}{2}\gamma_s+\frac{1}{2}\gamma_\sigma\right)\,r^3\cr
   F^{(1)}_{\;00}&=\frac{1}{3}\left[(g^{(1)}_{10}+f^{(1)}_{10})\,\gamma_s\gamma_\sigma
		+h^{(1)}_{10}\,(2+\gamma_s\gamma_\sigma)\right]\,r\cr
&\hskip12mm+\frac{2}{3}\left[(g^{(1)}_{12}+f^{(1)}_{12})\,\gamma_s\gamma_\sigma
		+h^{(1)}_{12}\,(1-\gamma_s\gamma_\sigma)\right]\,r
	+\frac{2}{5}g^{(1)}_{32}\,\left(\frac{1}{3}+\frac{2}{3}\gamma_s\gamma_\sigma\right)\,r^3\cr
\elb
So the amplitudes depend on eight parameters; $(g^{(1)}_{10}+f^{(1)}_{10})$, 
$g^{(1)}_{11}$, $g^{(1)}_{12}$, $g^{(1)}_{32}$, $f^{(1)}_{11}$, $f^{(1)}_{12}$,
$h^{(1)}_{10}$ and $h^{(1)}_{12}$.  It should be noted that the term proportional to
$h^{(1)}_{12}$ is purely relativistic, since $(1-\gamma_s\gamma_\sigma)$ vanishes 
in the non-relativistic limit $\gamma_s=\gamma_\sigma=1$.

   This example can also be applied to the case $1^-\to1^-+1^-$, 
where the two vector states are identical. 
We need to carry out the operation given in \eqn{a11x4}. 
The resulting amplitudes are
\blgb{12pt}{d6}
   F^{(1)}_{++}&=0, &
   F^{(1)}_{\;0+}&=\frac{1}{2}(g^{(1)}_{11}-f^{(1)}_{11})\,
   \left(\frac{1}{2}\gamma_s+\frac{1}{2}\,\gamma_\sigma\right)\,r
   +\frac{1}{4}(g^{(1)}_{12}-f^{(1)}_{12})\,
   \left(\gamma_s-\gamma_\sigma\right)\,r\cr
 F^{(1)}_{\;00}&=0, &   F^{(1)}_{\,+0}&=-\frac{1}{2}(g^{(1)}_{11}-f^{(1)}_{11})\,
   \left(\frac{1}{2}\gamma_s+\frac{1}{2}\,\gamma_\sigma\right)\,r
   +\frac{1}{4}(g^{(1)}_{12}-f^{(1)}_{12})\,
   \left(\gamma_s-\gamma_\sigma\right)\,r\cr
\elgb
The amplitudes depend on two parameters $(g^{(1)}_{11}-f^{(1)}_{11})$
and $(g^{(1)}_{12}-f^{(1)}_{12})$.  The terms corresponding to the second parameter, 
with $\ell+S=$ odd, vanish in the non-relativistic limit $\gamma_s=\gamma_\sigma=1$.

   The helicity-coupling amplitudes are, in the non-relativistic limit,
\blb{d5}
\sqrt{2}F^{(1)}_{++}
  &=\sqrt\frac{2}{3}G^{(1)}_{10}\,r-\sqrt\frac{2}{15}G^{(1)}_{12}\,r
  +\frac{1}{\sqrt{5}}G^{(1)}_{32}\,r^3\cr
\sqrt{2}F^{(1)}_{0+}
  &=\frac{1}{\sqrt{2}} G^{(1)}_{11}\,r
-\sqrt\frac{3}{10}G^{(1)}_{12}\,r-\frac{1}{\sqrt{5}}G^{(1)}_{32}\,r^3\cr
\sqrt{2}F^{(1)}_{+0}
  &=-\frac{1}{\sqrt{2}} G^{(1)}_{11}\,r
-\sqrt\frac{3}{10}G^{(1)}_{12}\,r-\frac{1}{\sqrt{5}}G^{(1)}_{32}\,r^3\cr
F^{(1)}_{00}
  &=-\frac{1}{\sqrt{3}} G^{(1)}_{10}\,r
  -\frac{2}{\sqrt{15}}G^{(1)}_{12}\,r+\sqrt\frac{2}{5}G^{(1)}_{32}\,r^3
\elb
In this case, there are four independent parameters, $G^{(1)}_{10}$, 
$G^{(1)}_{11}$, $G^{(1)}_{12}$, $G^{(1)}_{32}$.  
It is instructive to note that the normalization conditions given by \eql{a1x0}{a} and
\eql{a1x0}{b} are satisfied by the formula above.
Once again, we set $\gamma_s=\gamma_\sigma=1$
in \eqn{d4} and obtain
\blgb{24pt}{d7}
  G^{(1)}_{10}&=-\frac{1}{\sqrt{3}}\left(g^{(1)}_{10}+f^{(1)}_{10}+3h^{(1)}_{10}\right), &
  G^{(1)}_{11}&=g^{(1)}_{11}-f^{(1)}_{11}\cr
  G^{(1)}_{12}&=-\sqrt{\frac{5}{3}}\left(g^{(1)}_{12}+f^{(1)}_{12}\right), &
  G^{(1)}_{32}&=\sqrt{\frac{2}{5}}\;g^{(1)}_{32}
\elgb
which is consistent with \eqn{a12d}.
\sectn{Illustrative Examples III}{sc9}
\indnt
   We continue illustrating the decay amplitudes by considering the decay modes with
$J=s=\sigma=2$.  In order to make our paper concise and readable, we merely show
the allowed amplitudes in tabular forms and exhibit a few explicit examples of a given $J$,
$\ell$ and $S$ but we do not work out $F^J_{\lambda_s\lambda_\sigma}$; the reader
is invited to consult our C++ program posted online (see Section 10) for the full details.

   The decay amplitudes with $s+\sigma+\ell_J=$ even (odd) are given in Table IVa (Table IVb). 
\noindent\parbox[]{16cm}{
  \begin{center}
    Table IVa. \vtop{\hbox{\undl{Decay Amplitudes for $2\to 2+2$}}
	\vskip3pt\hbox{$R$ stands for the rank.}
		\vskip6pt}
  \end{center}
\begin{center}
  \begin{tabular}
{|@{\hspace{8pt}}r@{\hspace{12pt}}|@{\hspace{18pt}}l@{\hspace{18pt}}|@{\hspace{6pt}}
cc@{\hspace{8pt}}c@{\hspace{8pt}}c@{\hspace{8pt}}c@{\hspace{8pt}}cc@{\hspace{6pt}}|}
    \hline
    & $S$ & & $R(n_1,n_2)$ & & $R_\ell$ & & $R_J$ & \\
    \hline
    1&2             & &        $\onedot{4}(1,1)$ &   & 0 &       & 2 &    \\
    2&$0\;$---$\;4$ &    &$\twodots{4}(0,0)$        &   & 2 &$:$    & 2 &       \\
    3&$0\;$---$\;4$ &&$\onedot{4}(1,0)$ &$\cdot$ & 2 &$\cdot$& 2 &\\
    4&$0\;$---$\;4$& &$\onedot{4}(0,1)$ &$\cdot$ & 2 &$\cdot$& 2 &\\
    5&$0\;$---$\;4$&    &           4$(2,0)$&   $:$  & 2 &       & 2    &\\
    6&$0\;$---$\;4$&   &           4$(1,1)$        &   $:$
						    & 2     &       & 2    &\\
    7&$0\;$---$\;4$&    &           4$(0,2)$        &   $:$
 							   & 2     &       & 2    &\\
    8&$2,\;3,\;4$          &           &   $\onedot{4}(0,0)$        &   $:$
							    & 4     &$:$    & 2    &\\
    9&$2,\;3,\;4$&   &   ${4}(1,0)$&   $:\!\cdot$ & 4&$\cdot$    & 2    &\\
    10&$2,\;3,\;4$&   &   ${4}(0,1)$&   $:\!\cdot$ & 4&$\cdot$    & 2    &\\
    11&$4$   &    &   ${4}(0,0)$  &   $::$    & 6     &$:$    & 2    &\\
\hline
  \end{tabular}
\end{center}}
The decay amplitudes are, in the $J\,$RF,
\bln{a11w3a}
   \mbox{Row \ftm{0}1, Table IVa:}&\cr\qquad 
A^{(2)}_{0S}(s,\sigma;\delta)&={\Psi^{\alpha\nu}}_{\nu\tau}(1,1;S\delta)\;
		\,\tilde g^{\tau\rho}(w)\,\phi^*_{\alpha\rho}(2\delta)\cr
   \mbox{Row \ftm{0}7, Table IVa:}&\cr\qquad 
A^{(2)}_{2S}(s,\sigma;\delta)&=\Psi^{\alpha\beta\,\mu\nu}(0,2;S\delta)\;
	\chi_{\alpha\beta}(20)\,\phi^*_{\mu\nu}(2\delta)\cr
   \mbox{Row \ftm{0}9:, Table IVa:}&\cr\qquad 
A^{(2)}_{4S}(s,\sigma;\delta)&=\Psi^{\alpha\beta\,\mu\nu}(1,0;S\delta)\;
	\chi_{\beta\mu\nu\tau}(40)\,\tilde g^{\tau\rho}(w)\,\phi^*_{\rho\alpha}(2\delta)\cr
\eln
\noindent\parbox[]{16cm}{
  \begin{center}
    Table IVb. \vtop{\hbox{\undl{Decay Amplitudes for $2\to 2+2$}}
		\vskip3pt\hbox{$R$ stands for the rank.}
	\vskip6pt}
  \end{center}
\begin{center}
  \begin{tabular}
{|@{\hspace{8pt}}r@{\hspace{12pt}}|@{\hspace{18pt}}l@{\hspace{18pt}}|@{\hspace{6pt}}
cc@{\hspace{8pt}}c@{\hspace{8pt}}c@{\hspace{8pt}}c@{\hspace{8pt}}cc@{\hspace{6pt}}|}
    \hline
    & $S$ & & $R(n_1,n_2)$ & & $R_\ell$ & & $R_J$ & \\
    \hline
    1&1, 2, 3   &         &  $\onedot{4}(1',1')$   &  & 1     & $\cdot$ & $2'$ &          \\
    2&1, 2, 3   &      &  ${4}(2',2')$          &  & $1'$     &         & 2 &       \\
    3&1, 2, 3   &  &  ${4}(2',1')$    &$\cdot$ & 1     &         & $2'$ &   \\
    4&1, 2, 3   &         &  ${4}(1',2')$   &$\cdot$ & 1 & & $2'$    &     \\
    5&$1\;$---$\;4$&  &$\onedot{4}$$(1',1')$   && $3'$     &$:$          & 2    &         \\
    6&$1\;$---$\;4$&  &$\onedot{4}$$(1',0)  $   &$\cdot$& $3'$ &$\cdot$ & $2'$ &           \\
    7&$1\;$---$\;4$&  &$\onedot{4}$$(0,1')$   &$\cdot$& $3'$ &$\cdot$ & $2'$  &           \\
    8&$1\;$---$\;4$&  &${4}$$(1',1')$          &$:$    & 3 &$\cdot$ & $2'$    &           \\
    9&3, 4         &  &   4$(1',0)$   &$\cdot\!:$     & $5'$ &$\cdot$ & $2'$    &         \\
    10&3, 4&          &   4$(0,1')$   &$\cdot\!:$     & $5'$ &$\cdot$ & $2'$    &         \\
    11&3, 4&          &   4$(1',1')$   &$:$       & $5'$      &$:$     & 2    &           \\
\hline
  \end{tabular}
\end{center}}
and
\bln{a11w3b}
   \mbox{Row    \ftm{0}1, Table IVb:}&\cr\qquad 
A^{(2)}_{1S}(s,\sigma;\delta)&=\epsilon_{\alpha\beta\mu\nu}\,p^\alpha
	\Psi^{\beta\tau\,\rho\mu}(1',1';S\delta)\,\tilde g_{\tau\rho}(w)\;
	\chi_{\gamma}(10)\,\phi^{*\;\gamma\nu}(2'\delta)\cr
   \mbox{Row    \ftm{0}2, Table IVb:}&\cr\qquad 
A^{(2)}_{1S}(s,\sigma;\delta)&=\epsilon_{\alpha\beta\mu\nu}\,p^\alpha
	\Psi^{\beta\tau\,\rho\mu}(2',2';S\delta)\;
	\chi^{\nu}(1'0)\,\phi^*_{\tau\rho}(2\delta)\cr
   \mbox{Row     \ftm{0}8, Table IVb:}&\cr\qquad 
A^{(2)}_{3S}(s,\sigma;\delta)&=\epsilon_{\alpha\beta\mu\nu}\,p^\alpha
	\Psi^{\beta\tau\,\rho\mu}(1',1';S\delta)\;
	\chi_{\tau\rho\gamma}(30)\,\phi^{*\;\gamma\nu}(2'\delta)\cr
\eln
\sectn{Amplitudes for Sequential Decay}{sc10}
\indnt
   We now turn to a discussion of the decay amplitudes for which the daughter states
$s$ and $\sigma$ decay into, e.g. two or more pions.  To illustrate the techniques
necessary for such processes, it is sufficient to consider the decay $s\to\pi\pi$
and $\sigma\to3\pi$. So $s$ can represent any one of $f_0(600)$, $\rho(770)$ or  $f_2(1275)$,
while $\sigma$ could be $\omega(782)$, $a_1(1260)$, $a_2(1320)$ or $\pi_2(1670)$. 
The ``analyzer'' for the $\pi\pi$ system is the momentum
of one of the pions in its appropriate rest frame, 
whereas the analyzer for the $3\pi$ system has been chosen, for the purpose of illustration
in this section, as the normal to the decay plane in its rest frame.

   Consider the decay amplitude $J\to s+\sigma$ in the $J\,$RF as given in \eqn{a0}.
Let $(\vec x,\vec y,\vec z)$ stand for the coordinate system in the $J\,$RF, in which
the angles $(\theta,\phi)$ describe the orientation of $s$.
We go into the $s\,$RF and choose the coordinate axes $(\vec x_s,\vec y_s,\vec z_s)$ such that
\bln{q0}
\vec z_s\propto\vec s\,\big|_{\text{$J\,$RF}},
	\quad \vec y_s\propto\vec z\times\vec s\,\big|_{\text{$J\,$RF}}
	\quad\text{and}\quad \vec x_s=\vec y_s\times\vec z_s\,\big|_{\text{$s\,$RF}}
\eln
Here we have taken advantage of the fact that the direction of $s$ in the $J\,$RF is
preserved in the $s\,$RF, and the vector $\vec z\times\vec s$ is normal to the direction
of the Lorentz transformation and hence invariant under the transformation.
We note that this particular choice of the coordinate system in the $s\,$RF is dictated by
the choice of the arguments $(\phi,\theta,0)$ of the $D$-function 
which appears in \eqn{a0}\cite{SCh2}.  Let $(\theta_s,\phi_s)$ describe the orientation of one
of the pions for the decay $s\to\pi\pi$ in the $s\,$RF.  
Then, the decay amplitude is
\bln{q1}
   {\cal M}^s_\lambda(\theta_s,\phi_s)=\sqrt{\frac{2s+1}{4\pi}}\,F^s\,
		D^{s\cnj}_{\lambda\,0}(\phi_s,\theta_s,0)
\eln
where $F^s$ is the complex decay-coupling constant, which could be absorbed into $F^J_{\lambda\nu}$.
The description of the decay $\sigma\to3\pi$ proceeds along a similar path.
The coordinate axes $(\vec x_\sigma,\vec y_\sigma,\vec z_\sigma)$ in the $\sigma\,$RF are
\bln{q2}
\vec z_\sigma\propto\vec\sigma\,\big|_{\text{$J\,$RF}}=-\vec s\,\big|_{\text{$J\,$RF}},
	\quad \vec y_\sigma\propto\vec z\times\vec\sigma\,\big|_{\text{$J\,$RF}}
		=-\vec z\times\vec s\,\big|_{\text{$J\,$RF}}
	\quad\text{and}\quad \vec x_\sigma
	=\vec y_\sigma\times\vec z_\sigma\,\big|_{\text{$\sigma\,$RF}}
\eln
Let $(\theta_\sigma,\phi_\sigma)$ describe the orientation of the normal to the decay plane
for $\sigma\to3\pi$ in the $\sigma\,$RF, and let $\varphi_\sigma$ fix the direction of one of
the pions in the decay plane. Then, the decay amplitude is\cite{SCh2}
\bln{q3}
   {\cal M}^\sigma_\nu(\theta_\sigma,\phi_\sigma,\varphi_\sigma)=
	\sqrt{\frac{2\sigma+1}{8\pi^2}}\,\sum_\zeta F^\sigma_\zeta\,
		D^{\sigma\cnj}_{\nu\,\zeta}(\phi_\sigma,\theta_\sigma,\varphi_\sigma)
\eln
where $\zeta$ is the spin projection of $\sigma$ along the decay normal, 
i.e. $-\sigma\leq\zeta\leq+\sigma$, and hence
it is a rotationally invariant quantum number. $F^\sigma_\zeta$ is the complex decay-coupling
constant, which satisfies, from parity conservation in the decay,
\bln{q4}
	F^\sigma_\zeta=\eta_\sigma(-)^{\zeta+1}\,F^\sigma_\zeta
\eln
So, if $\eta_\sigma=+1(-1)$, we must have $F^\sigma_\zeta=0$ for $\zeta=$ even (odd).
Consider, for example, that $\sigma$ is the $\omega(782)$ and we examine its decay into
$\pi^+\pi^0\pi^-$.  Since $F^\sigma_\pm=0$, the only nonzero decay-coupling constant is
$F^\sigma_0$, which implies that the $D$-function with $\zeta=0$ is independent of $\varphi_\sigma$.
Integrate the resulting angular distribution over the angle; this is equivalent to multiplying
the decay amplitude \eqn{q3} by $\sqrt{2\pi}$ and setting $\varphi_\sigma=0$.  We see that
the amplitude is exactly the same as \eqn{q1} for $\rho\to\pi\pi$, a well-known result.

We are now ready to write down the full decay amplitude for $J\to s+\sigma$, $s\to\pi\pi$ and
$\sigma\to3\pi$.  It is simply the product of \eqn{a0}, \eqn{q1} and \eqn{q3}, i.e.
\bln{q5}
   {\cal M}^J_{\lambda\nu}(M;\theta,\phi)
	\;{\cal M}^s_\lambda(\theta_s,\phi_s)
	\;{\cal M}^\sigma_\nu(\theta_\sigma,\phi_\sigma,\varphi_\sigma)
\eln
where all the rotationally invariant quantum numbers appear either as super- or sub-scripts.
The overall amplitude depends on seven angles and on $M$ as arguments. Here the spin projection $M$
of $J$ in the $J\,$RF appears an argument in ${\cal M}$,
since it is {\em not} a rotationally invariant quantum number.  We note that the set of angles
$\{\theta_\sigma,\phi_\sigma,\varphi_\sigma\}$ are in reality the familiar Euler angles 
which fix the orientation of the $3\pi$ system in the $\sigma\,$RF.
\sectn{Conclusions and Discussions}{sc11}
\indnt
   We have given in this paper a general prescription for incorporating the functional
dependence of the Lorentz factors in a fully relativistic two-body decay amplitude.
We believe that our approach given in this paper is a proper and a natural way of
combining the helicity formalism of Jacob and Wick\cite{JW} 
with the tensor wave functions of 
Rarita and Schwinger\cite{RS}, Behrends and Fronsdal\cite{BrFr} and Zemach\cite{Zm}.
The central idea for such an approach has been presented in Section 6, and a number
of examples of the decay $J\to s+\sigma$ have been given in Section 7, 8 and 9.

   We need to elaborate on the measurability of the Lorentz factors worked out
in this paper.  For the purpose, we first point out that, for application to partial-wave analyses,
the amplitudes $A^J_{\ell S}$ must include not only the Breit-Wigner forms 
for the daughter states $s$ and $\sigma$ but also those for parent states $J$ (for the so-called
mass-independent global fits to the experimental data).  It is best to illustrate
the proposed formalism with two examples for the decay process:
\begin{center}\def\arraystretch{1.0}\begin{tabular}{|c|cc|c|c|}
\hline
$J\to s+\sigma$ & $w_0$ (MeV) & $\Gamma_0$ (MeV) & r (MeV/$c$)&Amplitudes\\
\hline	
$\pi_2(1670)\to\rho+\pi$ & 1672.4 & 259 & 2$\times$648&Section 7.2\\ 
$f_1(1285)\to\rho+\rho$ & 1281.8 & 24.1 & &Section 8.4\\ 
\hline     
\end{tabular}\end{center}

  Consider first the decay of the $\pi_2(1670)$.  The relative momentum $r\simeq$ 1.30 GeV
is large enough for $\gamma_s$ to be substantially different from one; it is a variable
which depend on the $\rho\pi$ effective mass $w$, as it ranges over, e.g. from
$w_0-\Gamma_0$ to $w_0+\Gamma_0$ and, in addition, it depends on the $\pi\pi$ effective mass
$m_{12}$ from the decay $\rho\to\pi+\pi$, designating the pions from the $\rho$ decay
by the subscripts 1 and 2.  The Lorentz factor is, in the $J\,$RF,
\bln{f0}
   \gamma_s=\frac{\sqrt{q^2+m_{12}^2}}{m_{12}}
\eln
where the effective mass $m_{12}$ is subject to the Breit-Wigner form for the $\rho$.
So the coefficient of $\gamma_s$ should be measurable, 
and hence add important new information on the decay property.  We advocate that
both approaches, non-relativistic [\eqn{a1}] and relativistic [\eqn{a10}], be used on the
data and assay the difference by comparing the fitted parameters using \eqn{a12g}.
Here the comparison should be carried out with the amplitudes at their maxima, i.e.
$w=w_{\rm th}$ is replaced by $w=w_0$ where $w_0$ is the mass of the $\pi_2(1670)$, and
$m_{12}$ of \eqn{f0} by $m_\rho$, the mass of the $\rho$.

   The situation with the second example above is different. 
We note that $w_0=1281.8$ MeV is less than $2w_\rho$.\footnote{Another example 
with a similar situation is the putative decay $\omega\to\rho+\pi$.} 
Let the first $\rho$ ($s$) decay into pions 1 and 2, while the
second $\rho$ ($\sigma$) into pions 3 and 4.  We need to set, in the $J\,$RF,
\bln{f1}
   \gamma_s=\frac{\sqrt{q^2+m_{12}^2}}{m_{12}},\quad
   \gamma_\sigma=\frac{\sqrt{q^2+m_{34}^2}}{m_{34}}
\eln
where $m_{12}$ and $m_{34}$ stand for the appropriate two-body effective masses.
They are constrained by the two Breit-Wigner forms for the $\rho$.
Because of the narrow width of the $f_1(1285)$, 
the $w$-mass dependence of the amplitudes is absent, i.e. for all practical purposes,
we could simply set $w=w_0$, where $w_0$ is the mass of the $f_1(1285)$.
The functional dependence of $\gamma_s$ and $\gamma_\sigma$
is important, owing to the finite width of the $\rho$, and it should be included in the analysis.
Once again, both approaches, non-relativistic [\eqn{a1}] and relativistic [\eqn{a10}], 
should be used in the analysis of data.  The resulting parameters can be compared
using the formula \eqn{a12d}, where the first requirement of 
$w=w_{\rm th}$ is replaced by $w=w_0$.  Here the second requirement $\gamma_s=\gamma_\sigma=1$ 
remains valid, since the Lorentz factors \eqn{f1} should be set at $m_{12}=m_{34}=w_0/2$
(so that $q=0$).

   A comment on the example of Section 8.1 is in order.  Consider the
hypothetical Higgs decay $H\to W^++W^-$.  As the Higgs mass increases to infinity,
we see that the helicity-coupling amplitudes $F^{(0)}_{\pm\pm}$ and $F^{(0)}_{00}$
become equally large [see \eqn{c92}], 
indicating that the vector character of the $W$ bosons is
preserved in this limit.  This is in contrast to the well-known equivalence theorem\cite{Higgs},
which states that, in the infinite mass limit of the Higgs particle, the $W$ bosons
behave like scalars (the Goldstone bosons).  This phenomenon can be traced to the
interactions of the $W$ bosons with the scalar field, with the result that the final
$W$ bosons contain both spin-1 and spin-0 components in the Standard Model.
It is clear from this point of view that our helicity-coupling amplitudes have nothing
to do with the interacting $W$ bosons, and hence our phenomenological model does not
apply to the hypothetical Higgs decay $H\to W^++W^-$ in the large-mass limit.

  The Zemach amplitudes\cite{Zm} are commonly used used in partial-wave analyses.
They correspond to the non-relativistic limit, i.e.  $\gamma_s=1$ and $\gamma_\sigma=1$,
of the amplitudes worked out in this paper.  However, we need to caution the reader that
we do not imply that the Zemach amplitudes are {\em correct} only in the non-relativistic limit.
We merely wish to emphasize that our decay amplitudes with the Lorentz factors are
an equally valid approach to writing down the amplitudes; our approach is  more general,
only in the sense that our decay amplitudes lead to the Zemach amplitudes
when we set $\gamma_s=1$ and $\gamma_\sigma=1$.

 In our formulation of the decay amplitudes, the orbital angular momentum $\ell$
leads naturally to the barrier factor $r^\ell$ in the formula.  
Since the factor is necessary to ensure 
that the amplitudes are singularity-free at the threshold, i.e. $r\to0$, we need to
``damp'' it in the limit $r\to\infty$.  This is accomplished by substituting $r^\ell$
with the Blatt-Weisskopf (BW) barrier factors\cite{BW}.  
The practitioners of partial-wave analyses in fact use the BW factors in their 
Zemach amplitudes.  Let $f_\ell(u)$ be the BW factor where $u=r/r_0$ and
$r_0=0.1973$ GeV/$c$. Here $r_0$ corresponds to a sphere of radius 1 fermi for the strong
interaction responsible for the decay in the coordinate space.  It can be shown that
\bln{f2}
   f_\ell(u)\Big|_{u\to0}=\frac{2^\ell\,\ell!}{(2\ell)!}\,u^\ell,\qquad
   f_\ell(u)\Big|_{u\to\infty}=1	
\eln
which show that $f_\ell(u)$ is indeed the BW factor with the appropriate expected properties.
We give below the explicit expressions of the BW factor for $\ell=0\to5$:
\bln{f2a}
    f_0(u)&=1,\quad f_1(u)=u\big[1+u^2\big]^{-1/2},\quad
    		f_2(u)=u^2\big[(u^2-3)^2+9u^2\big]^{-1/2},\cr
    	f_3(u)&=u^3\big[9(2u^2-5)^2+u^2(u^2-15)^2\big]^{-1/2}\cr
    	f_4(u)&=u^4\big[(u^4-45u^2+105)^2+25u^2(2u^2-21)^2\big]^{-1/2}\cr
    	f_5(u)&=u^5\big[225(u^4-28u^2+63)^2+u^2(u^4-105u^2+945)^2\big]^{-1/2}\cr
\eln
The helicity-coupling amplitude $F$ as given in \eqn{a10} involves a factor
$A^J_{\ell S}(s,\sigma;\delta)\,r^\ell$ which must now be replaced by
\bln{f3}
   A^J_{\ell S}(s,\sigma;\delta)\to
      \bigg[\,\left(\frac{p}{i\,w_{\rm th}}\right)^{n_0},\; \psi(s,\sigma;S\delta),\;
			\chi(\ell0),\;\phi^*(J\delta)\,\bigg]_w\quad\mbox{and}\quad
   r^\ell\to f_\ell\left(\frac{r}{r_0}\right)
\eln
where the first factor has already been given in \eqn{a11a}.
The right-hand side of the equations above are given entirely in unitless quantities,
and they give the non-relativistic limit shown in \eqn{a12d} and \eqn{a12g}.

   We have presented several additional decay amplitudes in Appendix C,
which lie outside the scope of the prescription for constructing covariant amplitudes 
as given in Section 6.  The main reason for not including these amplitudes in the main
text of this paper is because they violate the rule that an amplitude with $\ell$ must
have a dependence $r^\ell$.  If for some dynamical reason such an amplitude is needed,
then we must keep in mind that the extra factors $r$ result from the time-components
of the wave functions for $s$ and $\sigma$ [see \eqn{polz}].  And hence they are fundamentally
different from the $r^\ell$ dependence, which after all comes from a need to write down
singularity-free amplitudes as $r\to0$.  As a result the procedure of substituting $r^\ell$ by
a BW factor does not extend to the extra $r$ factors which are given in Appendix C.

   It is clear, in retrospect, that the result of an earlier paper\cite{SCh0} by one of us
applies only to the situation in which one or both of the decay products are massless
particles, e.g. photons.  In the paper it has been shown that the functions involving the Lorentz
factors can expressed succinctly in a closed form.  Unfortunately, such a closed expression
seems impractical to us, and so we resort to the calculations on a case-by-case basis.
It is hoped that the illustrative examples given in Sections 7, 8 and 9 are sufficiently diverse
to give the reader how he should go about calculating the Lorentz factors for his own case.
We have, in addition, posted a general C++ program which gives the helicity-coupling amplitudes
for any integer values of $J$, $s$ and $\sigma$, on two websites http://cern.ch/suchung
and http://cern.ch/friedric.
We have checked that all the results presented in the main text agree with the C++ program.
The reader is free to download for his own use.  Any comments and/or queries on the program
should be addressed to JF.
\appendx
\section*{Appendix A: \vtop{\hbox{The Lorentz Group}\vskip-10pt 
					\hbox{in Four-Momentum Space}}}
\setcounter{equation}{0}
\def\theequation{A.\arabic{equation}}
\indnt
   The homogeneous Lorentz transformations are defined through
\bln{p0}
   p^{\prime\,\mu}={\Lambda^\mu}_\nu\,p^\nu\quad{\rm and}\quad 
	g_{\mu\nu}\,{\Lambda^\mu}_\rho\,{\Lambda^\nu}_\tau=g_{\rho\tau}
\eln
Our Lorentz metric $g^{\mu\nu}=g_{\mu\nu}$ has signature $(+,-,-,-)$.
\bln{p0a}
   p^\mu&=(E,p^1,p^2,p^3)=(E,p_x,p_y,p_z)\cr
   p_\mu&=g_{\mu\nu}\,p^\nu=(E,p_1,p_2,p_3)=(E,-p_x,-p_y,-p_z)\cr
\eln
Let $w$ be the mass associated with $p$ and adopt a notation in which
$p$ indicates {\em both} the four-momentum and the {\em magnitude} of the
3-momentum, i.e. 
\bln{p0b}
  E^2=w^2+p^2,\quad p^2=p^2_x+p^2_y+p^2_z
\eln
For each Lorentz transformation $\Lambda$, there exists a `generalized angular momentum' 
operator $J^{\mu\nu}$ given by
\bln{h0}
   \Lambda
   =\exp[-\frac{i}{2}\,\omega_{\mu\nu}\,J^{\mu\nu}\,]
\eln
where $\omega_{\mu\nu}$ is an antisymmetric matrix whose elements correspond to the six
independent parameters of the Lorentz group.
$J^{\mu\nu}$ is an antisymmetric operator imbedded in the four-momentum space.
It can be shown\cite{MG} that
\bln{h2}
   {\left(J^{\mu\nu}\right)^\rho}_\sigma
	=i\,\left(g^{\mu\rho}\,{\delta^\nu}_\sigma-g^{\nu\rho}\,{\delta^\mu}_\sigma\right)
\eln
We define the angular momentum and the boost operator for $n=1$, 2 or 3 via
\bln{h3}
   J^n&=\frac{1}{2}\,\varepsilon^{njk}\,J^{jk},\quad K^n=J^{0n}=-J^{n0}\cr
\eln
where $\varepsilon^{123}=+1$.  Explicitly, they can be expressed as
\bln{h4}
  {\left(J^n\right)^\rho}_\sigma&=0,\quad\mbox{except}\quad 
	{\left(J^n\right)^j}_k=-i\,\varepsilon_{njk}\cr
	{\left(K^n\right)^\rho}_\sigma&=i\left(\delta_{\rho0}\delta_{n\sigma}+\delta_{\rho n}\delta_{0\sigma}\right)
\eln
and they satisfy the usual commutation relations
\bln{h31}
   [J^i,J^j]=i\,\varepsilon^{ijk}\,J^k,\quad
   [J^i,K^j]=i\,\varepsilon^{ijk}\,K^k,\quad
   [K^i,K^j]=-i\,\varepsilon^{ijk}\,J^k
\eln

   The spin-1 wave functions {\em at rest} are
\bln{h3c}
e^\mu(0)=\pmtrx{0\cr0\cr0\cr1\cr},\qquad
e^\mu(\pm1)=\mp\frac{1}{\sqrt{2}} \pmtrx{0\cr1\cr\pm i\cr0\cr}
\eln
It can be shown that, with $J_x=J^1$, $J_y=J^2$ and $J_z=J^3$, 
\blb{h3d}
  J^2\,e(m)&=j(j+1)\,e(m),\quad j=1,\quad m=-1,\ 0,\ +1\cr
  J_z\,e(m)&=m\,e(m),\quad m=-1,\ 0,\ +1\cr
  J_\pm\,e(0)&=\sqrt{2}\,e(\pm1),\quad J_\pm\,e(\mp1)=\sqrt{2}\,e(0),\quad J_\pm\,e(\pm1)=0\cr
\elb
where, with $J_x=J^1$, $J_y=J^2$ and $J_z=J^3$,  
$J^2=J^2_x+J^2_y+J^2_z$ and $J_\pm=J_x\pm i\,J_y$.  Consider a boost along the $z$-axis
which takes the rest-state wave functions $e(m)$ to $e(\vct{p},m)$, where $\vct{p}$
has only one nonzero component, i.e. the $z$-component.  
Again, with the notation $K_x=K^1$, $K_y=K^2$ and $K_z=K^3$, we can write
$B_z(p)=\exp[-i\,\alpha\,K_z]$ and find
\bln{h5}
{[B_z(p)]^\rho}_\sigma&=\pmtrx{\cosh{\alpha}&0&0&\sinh{\alpha}\cr
          0&1&0&0\cr
         0&0&1&0\cr
   \sinh{\alpha}&0&0&\cosh{\alpha}}
,\, 
{[B^{-1}_z(p)]^\rho}_\sigma=\pmtrx{\cosh{\alpha}&0&0&-\sinh{\alpha}\cr
          0&1&0&0\cr
         0&0&1&0\cr
   -\sinh{\alpha}&0&0&\cosh{\alpha}}
\eln
where $\cosh\alpha=E/w$ and $\sinh\alpha=p/w$.
We see that
\bln{h6}
e(\vct{p},m)=B_z(p)\,e(m),\qquad e^\mu(\vct{p},0)=\pmtrx{\eta\cr0\cr0\cr\gamma\cr},
	\qquad e^\mu(\vct{p},\pm1)=\mp\frac{1}{\sqrt{2}} \pmtrx{0\cr1\cr\pm i\cr0\cr}
\eln
where $\eta=p/w=\sinh\alpha=\gamma\beta$ and 
$\gamma=E/w=\cosh\alpha$ is the so-called Lorentz factor. The boosted wave functions
$e(\vct{p},m)$ satisfy the transversality condition 
\bln{h7}
p_\mu\,e^\mu(\vct{p},m)=0
\eln

   The relativistic spin $W^\mu(p)$, operating on the states with an eigenvalue of $p^\mu$
is
\bln{h8}
   W^\mu(p)=\frac{1}{2}\,\varepsilon^{\,\mu\,\alpha\,\beta\,\gamma}\,
	{p_\alpha}\,J_{\beta\gamma}
	=\frac{1}{2}\,\varepsilon^{\,\mu\,\alpha\,\beta\,\gamma}\,
	J_{\alpha\beta}\;p_\gamma
\eln
where we use the definition $\varepsilon_{0123}=+1$ so that $\varepsilon^{0123}=-1$.
Here $p_\mu$ is {\em not} an operator.  We see that
\bln{h8a}
   W^0(p)&=\vct{J}\cdot\vct{p}\cr
   \vct{W}(p)&=E\,\vct{J}+\vct{K}\times\vct{p}\cr
\eln
It can be shown that
\bln{h9}
   -W_\mu(p)W^\mu(p)=j(j+1)w^2,\quad j=1
\eln
appropriate for particles with spin 1.  We are now ready to define the `total intrinsic
spin' operator $S^n(p)$ via
\bln{h10}
   w\,S^n(p)=B_z(p)\,W^n(p)\,B_z^{-1}(p)
\eln
confining ourselves to boosts along the $z$-axis only.
It is clear by the definition that the actions of $wS^n(p)$ on $e(\vct{p},m)$
are exactly the same as those of $wJ$ on the at-rest states $e(m)$.
And they obey the Lie algebra of angular momentum
\bln{h10a}
   [S^i(p),S^j(p)]=i\,\epsilon^{ijk}\,S^k(p)
\eln
Once again define $S^1(p)=S_x(p)$, $S^2(p)=S_y(p)$ and $S^3(p)=S_z(p)$.  It can be shown that
\bln{h11}
   w\,S_z(p)=w\,J_z,\quad w\,S_x(p)=E\,J_x+pK_y,\quad w\,S_y(p)=E\,J_y-pK_x
\eln
With
\bln{h11a}
w^2\,S^2(p)&=w^2\left[S^2_x(p)+S^2_y(p)+S^2_z(p)\right]
\eln
we find
\bqx{h11b}
   w\,S_x(p)&=i\,\left(\mtrx{
0 & 0 & +p & 0\cr
0 & 0 & 0 & 0\cr
+p & 0 & 0 & -E\cr
0 & 0 & +E & 0\cr
}\right),\quad
   w\,S_y(p)=i\,\left(\mtrx{
0 & -p & 0 & 0\cr
-p & 0 & 0 & +E\cr
0 & 0 & 0 & 0\cr
0 & -E & 0 & 0\cr
}\right)&\mylbl{a}\cr\cr
w\,S_z(p)&=w\,\left(\mtrx{
0 & 0 & 0 & 0\cr
0 & 0 & -i & 0\cr
0 & +i & 0 & 0\cr
0 & 0 & 0 & 0\cr
}\right),\quad   w^2\,S^2(p)=2\,\left(\mtrx{
-p^2 & 0 & 0 & E\,p\cr
0 & w^2 & 0 & 0\cr
0 & 0 & w^2 & 0\cr
-E\,p & 0 & 0 & E^2\cr
}\right)&\mylbl{b}\cr
\eqx
Taking the usual raising and lowering operators $S_\pm(p)=S_x(p)\pm i\,S_y(p)$,
one can show that the boosted spin-1 wave functions $e(\vct{p},m)$ given by \eqn{h6} constitute
the standard representation for the angular momentum operator $\vec S(p)$ defined above.
To be more specific,
the $4\times4$ matrices derived from \eql{h11b}{a} and \eql{h11b}{b} acting on
the 4-dimensional eigenvectors $e(\vct{p},m)$ satisfy the familiar relationships for 
the operators $\{J_\pm$, $J_z$ and $J^2\}$ acting on $|j\,m\ket$ with $j=1$ [ see \eqn{h3d}].
   
   Define an arbitrary rotation by the Euler angles $\{\alpha,\beta,\gamma\}$
\bln{h12}
   R^S(\alpha,\beta,\gamma)
	=\exp[-i\,\alpha\,S_z(p)]\,\exp[-i\,\beta\,S_y(p)]\exp[-i\,\gamma\,S_z(p)]
\eln
We see that
\bln{h13}
   R^S(\alpha,\beta,\gamma)\,e^\mu(\vct{p},m)=\sum_{m'}\,e^\mu(\vct{p},m')\,
	D^{(1)}_{m'\,m}(\alpha,\beta,\gamma)
\eln
Note that the vector $\vct{p}$, defined to be along the $z$-axis, remain {\em invariant}
through the rotation.  Consider now a two-body system $(1+2)$ 
in a state of total intrinsic spin $S$,
where the two-body $\vec S(p)$ is defined through
\bln{h13a}
   S_a(p)=S^{(1)}_a(p)+S^{(2)}_a(p), \quad a=\{x,y,z\}
\eln
and the two-body wave function in a state of total intrinsic spin $S$ is
\bln{h14}
   e^{\mu\nu}(\vct{p},Sm)=\sum_{m_1\,m_2}(1m_1\,1m_2|Sm)\,
	e^\mu(\vct{p_1},m_1)\,e^\nu(\vct{p_2},m_2)
\eln
where $S=0$, 1 or 2 and $\vct{p}=\vct{p_1}+\vct{p_2}$.  We find
\bln{h15}
   R^S(\alpha,\beta,\gamma)\,e^{\mu\nu}(\vct{p},Sm)=\sum_{m'}\,e^{\mu\nu}(\vct{p},Sm')\
		D^S_{m'\,m}(\alpha,\beta,\gamma)
\eln
Note again that $\vct{p}$ is not affected by the rotation.  In contrast,
the orbital angular momentum operator $\vct{L}=\vct{J}-\vct{S}$ act only on $\vct{p}$
and does {\em not} affect the spin part of the wave function.

We have thus succeeded in separating out the total intrinsic part
from the orbital angular momentum part in a two-body system.
\section*{Appendix B: Spherical Harmonics}
\setcounter{equation}{0}
\def\theequation{B.\arabic{equation}}
\indnt
Consider the breakup momentum $\vec r$ in the  $J$\,RF,
whose direction is given by $\Omega=(\theta,\phi)$. A rank-$\ell$ tensor formed out of
a single vector $r_\alpha$ can be contracted
with the rank-$\ell$ tensor $\phi(\ell\,m)$ of \eqn{s0}. The resulting scalar is
a function of $\Omega$ with quantum numbers $\ell$ and $m$; so it must be proportional
to the spherical harmonics $Y^m_\ell(\Omega)$:
\bln{cx0}
   r^\ell\;Y^m_\ell(\Omega)=\sqrt{\frac{2\ell+1}{4\pi}}\;\left(c^{-1}_\ell\right)\;
&\big\{r_\alpha\,r_\beta\,r_\gamma\ldots \big\}\cr
&\hskip18pt\times\big\{\tilde g^{\alpha\,\alpha'}(w)\,\tilde g^{\beta\beta'}(w)\,
				\tilde g^{\gamma\gamma'}(w)\ldots\big\}\cr
&\hskip48pt\times\big\{\phi_{\alpha'\,\beta'\,\gamma'\ldots}(\ell\,m)\big\}
\eln
where $c_\ell$ is that given by \eqn{s9}.  (We show at the end of this appendix that
the proportionality constant is indeed given by $(c_\ell)^{-1}$.)
Define, in the $J$\,RF,
\blb{cx1}
   r\,\tau(\pm)&=\vec r\cdot\vec\phi(\pm)
	=\mp\frac{r}{\sqrt{2}}\;{\rm e}^{\,\pm i\,\phi}\,\sin\theta\cr
   r\,\tau(0)&=\vec r\cdot\vec\phi(0)= r\,\cos\theta
\elb
so that
\bln{cx2}
   Y^m_\ell(\Omega)=\sqrt{\frac{2\ell+1}{4\pi}}\;
	\left[\frac{(2\ell)!}{2^\ell\,(\ell!)^2}\right]^{1/2}\!
	\phi\Big\{\ell\,m;\;\phi_\alpha(\nu)\to\tau(\nu)\Big\},\quad
		\nu=-1,\;0,\;+1
\eln
Here $\phi(\ell\,m)$ is a not a tensor but a scalar in which each of the arguments is
replaced by a scalar $\tau(m)$ (not a vector) given in \eqn{cx1}.
It is helpful to illustrate this with an example.  The spherical harmonics for $\ell=1$ are 
readily obtained from \eqn{cx1}.  For $\ell=0$ and $\ell=2$, the general wave functions of \eqn{s0},
or more specifically the rank-2 tensor for spin-two
wave functions given in \eqn{s3}, can be combined with \eqn{cx2}, to find
\blb{cx3}
   Y^0_0(\Omega)&=\sqrt{\frac{1}{4\pi}},\quad
	Y^1_1(\Omega)=\sqrt{\frac{3}{4\pi}}\;\tau(+)
	=-\sqrt{\frac{3}{8\pi}}\;{\rm e}^{\,i\,\phi}\,\sin\theta\cr
   Y^1_0(\Omega)&=\sqrt{\frac{3}{4\pi}}\;\tau(0)
	=\sqrt{\frac{3}{4\pi}}\;\cos\theta\cr
   Y^2_2(\Omega)&=\sqrt{\frac{5}{4\pi}}\;\sqrt{\frac{3}{2}}\;\tau(+)\tau(+)
	=\frac{1}{4}\sqrt{\frac{15}{2\pi}}\;{\rm e}^{\,2\,i\,\phi}\,\sin^2\theta\cr
   Y^1_2(\Omega)&=\sqrt{\frac{5}{4\pi}}\;\sqrt{\frac{3}{2}}\;
	\frac{1}{\sqrt{2}}\big[\tau(+)\tau(0)+\tau(0)\tau(+)\big]
	=-\sqrt{\frac{15}{8\pi}}\;{\rm e}^{\,i\,\phi}\,\sin\theta\,\cos\theta\cr
   Y^0_2(\Omega)&=\sqrt{\frac{5}{4\pi}}\;\sqrt{\frac{3}{2}}\;
	\frac{1}{\sqrt{6}}\big[\tau(+)\tau(-)+\tau(-)\tau(+)+2\tau(0)\tau(0)\big]\cr
	&=\sqrt{\frac{5}{4\pi}}\;\left(\frac{3}{2}\cos^2\theta-\frac{1}{2}\right)
\elb
One can show that the formula holds for other values of $\ell$.

   Define two real functions of $\theta$ by dropping the $\phi$ dependence in \eqn{cx1}
\bln{cx4}
   \tau_0(\pm)=\mp\frac{1}{\sqrt{2}}\;\sin\theta,\quad
   \tau_0(0)=\cos\theta
\eln
Noting that
\bln{cx5}
   d^{\;\ell}_{m\,0}(\theta)=\sqrt{\frac{4\pi}{2\ell+1}}\,Y^m_\ell(\Omega)\,{\rm e}^{-i\,m\,\phi}
\eln
we obtain, from \eqn{cx2},
\bln{cx6}
   d^{\;\ell}_{m\,0}(\theta)=
	\left[\frac{(2\ell)!}{2^\ell\,(\ell!)^2}\right]^{1/2}\!
	\phi\Big\{\ell\,m;\;\phi_\alpha(\nu)\to\tau_0(\nu)\Big\},\quad
		\nu=-1,\;0,\;+1
\eln
For example, we find that $d^{\;\ell}_{m\,0}(\theta)$'s for $\ell=3$ are, from \eqn{s4},
\blb{cx7}
   d^{\;3}_{3\,0}(\theta)&=\sqrt{\frac{5}{2}}\;\tau_0(+)\,\tau_0(+)\,\tau_0(+)
	=-\frac{\sqrt{5}}{4}\;\sin^3\theta\cr
   d^{\;3}_{2\,0}(\theta)&=\sqrt{\frac{15}{2}}\;\tau_0(+)\,\tau_0(+)\,\tau_0(0)
	=\frac{1}{2}\sqrt{\frac{15}{2}}\;\sin^2\theta\,\cos\theta\cr
   d^{\;3}_{1\,0}(\theta)&=\sqrt{\frac{3}{2}}\;
	\big[\tau_0(+)\,\tau_0(+)\,\tau_0(-)+2\,\tau_0(+)\,\tau_0(0)\,\tau_0(0)\big]\cr
	&=-\sqrt{3}\;\sin\theta\,\left(\frac{5}{4}\cos^2\theta-\frac{1}{4}\right)\cr
   d^{\;3}_{0\,0}(\theta)
	&=3\,\tau_0(+)\,\tau_0(-)\,\tau_0(0)+\tau_0(0)\,\tau_0(0)\,\tau_0(0)\cr
	&=\left(\frac{5}{2}\,\cos^2\theta-\frac{3}{2}\right)\cos\theta
\elb

   In order to show that the proportionality constant is indeed $(c_\ell)^{-1}$, 
consider two special cases. 
First, set $\theta=0$. From the definition of the
$d$-functions, 
\bln{k2}
	d^\ell_{m0}(\theta)=\bra\ell\,m|R_y(\theta)|\ell\,0\ket
\eln
where $R_y(\theta)$ is a rotation by $\theta$ around the $y$-axis,
we must have $d^\ell_{m\,0}(0)=1$ for $m=0$ and $d^\ell_{m\,0}(0)=0$ for $m\neq0$.
Since $\tau_0(\pm)=0$ and $\tau_0(0)=1$ for $\theta=0$, we find
\bln{k3}
  \phi\Big\{\ell\,0;\;\phi_\alpha(\nu)\to\tau(\nu)\Big\}_{\theta=0}
	  &=\left[\frac{2^\ell\,(\ell!)^2}{(2\ell)!}\right]^{1/2}
\eln
We see that \eqn{cx6} is satisfied.  Consider next the case of $\theta=\pi/2$ and $m=\ell$.
For the purpose, we write down
the $d_{m0}(\theta)$-functions as given by Rose\cite{Rose},
\bln{k4}
 d^\ell_{m0}(\theta)&=(-)^{m}(\ell!)\,
         [(\ell+m)!(\ell-m)!]^{1/2}
  	\sum^{k_2}_{k=k_1}
        \frac{(-)^k\cos^{2\ell-n}(\theta/2)\sin^n(\theta/2)}{(\ell-m-k)!(\ell-k)!(m+k)!k!}\cr
\eln
where $n=m+2k$ and $k$ is a non-negative integer ranging between
$k_1$ and $k_2$ given by
\bln{k5}
   k_1=0, \quad\quad k_2=\min\{\ell-m,\ell\}
\eln
Since $m=\ell$, we see that $k_1=k_2=0$.  So the sum in \eqn{k4} reduces to just one term with
$k=0$.  We find, noting that $\sin(\theta/2)=\cos(\theta/2)=1/\sqrt{2}$,
\bln{k6}
d^\ell_{\ell 0}(\pi/2)&=(-)^\ell\,\left[\frac{(2\ell)!}{2^{2\ell}\,(\ell!)^2}\right]^{1/2}
\eln
On the other hand, we obtain, noting that $\sin(\pi/2)=1$,
\bln{k7}
\phi\Big\{\ell\,\ell;\;\phi_\alpha(\nu)\to\tau(\nu)\Big\}_{\theta=\pi/2}
	=\underbrace{\tau_0(+)\,\tau_0(+)\cdots}_{\ell}\bigg|_{\theta=\pi/2}
		=(-)^\ell\left(\frac{1}{\sqrt{2}}\right)^\ell
\eln
Here again we see that the proportionality constant given in \eqn{cx6} is satisfied.

   We have worked out the spherical harmonics for $\ell=$0, 1 and 2; and the $d$-functions
for $\ell=3$, which have been checked with standard references 
on such mathematical functions.  We have, in addition, shown that \eqn{cx6} is true
for two special cases of arbitrary integer $\ell$.
We have thus shown that \eqn{cx6} must be true in general, with the proportionality constant
being independent of both $\theta$ and $m$.
\section*{Appendix C: Additional Decay Amplitudes}
\setcounter{equation}{0}
\def\theequation{C.\arabic{equation}}
The purpose of this appendix is explore a few decay amplitudes $A$ not included in the main text.
They are exluded from our list of allowed amplitudes, because they induce in $A$ an additional dependence on $r$,
which are exluded according to the $r^\ell$ rule as given in the prescriptions following \eqn{a11}.
Consider the decay $J\to s+\sigma$,
where $J=1,2$, $s=1,2$ and $\sigma=0$ or 1.  We consider $\ell=1$ only.
In the vector notation, we have $\vec J=\vec S+\vec\ell$ and $\vec S=\vec s+\vec\sigma$.

The first case we wish to work out is for $\sigma=0$, i.e.
\bln{aw0}
	1^-\to1^-+0^-
\eln
The decay amplitude we want to consider here is, with $J=1$, $S=s=1$ and $\ell=1$,
\bln{aw1}
   A^J_{\ell\,s}(s\delta)=r_{\alpha}\;\omega^{\alpha}(s\,\delta)\;
	\chi^\beta(\ell\,0)\,\phi^*_{\beta}(J\,\delta)
\eln
In the $J$RF, the decay amplitudes are
\bln{aw1a}
   A^J_{\ell\,s}(s\delta)&=0,\quad\delta=\pm1\cr
	&=\left(\frac{W}{2m}\right)\,r,\quad\delta=0\cr
\eln
We do not include this decay amplitude because it apparently violates the parity conservation, 
since the full decay amplitude $F^J$ should be proportional to $r$ for $\ell=1$ and {\em not} $r^2$.
The second case we examine for $\sigma=0$ is
\bln{aw2}
	1^+\to2^++0^-
\eln
with its decay amplitude given by
\bln{aw3}
   A^J_{\ell\,s}(s\delta)=p_{\alpha}\,p_{\beta}\,\omega^{\alpha\beta}(s\,\delta)\;
	\chi^\gamma(\ell\,0)\,\phi^*_{\gamma}(J\,\delta)
\eln
In the $J$RF, the decay amplitudes are
\bln{aw3a}
   A^J_{\ell\,s}(s\delta)&=0,\quad\delta=\pm1\cr
	&=\frac{1}{2\sqrt{6}}\,\left(\frac{W^2}{m\,\mu}\right)\,r^2,\quad\delta=0\cr
\eln
This does not violate parity but the full decay amplitude $F^J$ is now proportional to $r^3$
for an $\ell=1$ decay mode.  For this reason, we do not allow such a decay amplitude.

   We now go over to the case $\sigma=1$ and consider the decay
\bln{ax0}
	1^+\to1^++1^-
\eln
We wish to work out the following invariant amplitude
\bln{ax0a}
   A^J_{\ell\,S}(S\delta)=\epsilon_{\mu\,\alpha\,\beta\,\gamma}\;\psi^{\mu\alpha}(S\,\delta)\,
	\chi^\beta(\ell\,0)\,{\phi}^{*\,\gamma}(J\,\delta)
\eln
where $\epsilon_{0123}=+1$ (so that $\epsilon^{0123}=-1$) and
\bln{ax1}
   \psi^{\mu\alpha}(S\,\delta)=\sum_{m_a\,m_b}\,(1m_a\;1m_b|S\delta)\;
	\omega^\mu(s\,m_a)\,\varepsilon^\alpha(\sigma\,m_b),\quad\delta=m_a+m_b
\eln

   Introducing a shorthand notation
\bln{ax2}
   \big[ a\,b\,c\big]=\epsilon_{ijk}\,a^i\,b^j\,c^k
	=(\,\vec a\cdot\vec b\times\vec c\;)=(\,\vec a\times\vec b\cdot\vec c\;)
\eln
so that
\bln{ax2a}
  \big[ a\,b\,c\,d\big]\equiv\epsilon_{\mu\,\alpha\,\beta\,\gamma}\;
			a^\mu\,b^\alpha\,c^\gamma\,d^\delta
=a^0\,\big[b\,c\,d\big]-b^0\,\big[a\,c\,d\big]+c^0\,\big[a\,b\,d\big]-d^0\,\big[a\,b\,c\big]
\eln
and we can write, in the $J$RF,
\bln{ax3}
   A^J_{\ell\,S}(S\delta)&=\sum_{m_a\,m_b}\,(1m_a\;1m_b|S\delta)\;
	\epsilon_{\mu\,\alpha\,\beta\,\gamma}\,	
	\omega^\mu(s\,m_a)\,\varepsilon^\alpha(\sigma\,m_b)\,
	\chi^\beta(\ell\,0)\,\phi^{*\,\gamma}(J\,\delta)\cr
    &=\sum_{m_a\,m_b}\,(1m_a\;1m_b|S\delta)\cr
	&\hskip60pt\times\Bigg\{\omega^0(s\,m_a)
		\Big[\varepsilon(\sigma\,m_b)\,\chi(\ell\,0)\,{\phi^*}(J\,\delta)\Big]\cr
	&\hskip140pt-\Big[\omega(s\,m_a)\,\chi(\ell\,0)\,{\phi^*}(J\,\delta)\Big]
					\varepsilon^0(\sigma\,m_b)\Bigg\}\cr
\eln
since $\chi^0=0$ and $\phi^0=0$ in the $J$\,RF.

   This leads to, dropping $J$, $s$ and $\sigma$ for compact notation,
\bln{ax4}
   A^J_{\ell\,S}(S0)&=0\cr
   A^J_{\ell\,S}(S+)&=(10\;11|S1)\,\omega^0(0)\Big[\varepsilon(+)\chi(0)\phi^*(+)\Big]
	-(11\;10|S1)\,\Big[\omega(+)\chi(0)\phi^*(+)\Big]\varepsilon^0(0)\cr
	&=(i)\Big[(10\;11|S1)\,\omega^0(0)-(11\;10|S1)\,\varepsilon^0(0)\Big]\cr
	&=(i)\Big[(10\;11|S1)\,\gamma_s\beta_s+(11\;10|S1)\,\gamma_\sigma\beta_\sigma\Big]\cr
	&=0,\quad S=0\cr
	&=\frac{i}{\sqrt{2}}
		\big(-\gamma_s\beta_s+\gamma_\sigma\beta_\sigma\;\big)
		=\frac{i\,r}{2\sqrt{2}}\left(-\frac{1}{m}+\frac{1}{\mu}\right),\quad S=1\cr
	&=\frac{i}{\sqrt{2}}
		\big(\;\gamma_s\beta_s+\gamma_\sigma\beta_\sigma\;\big)
	=\frac{i\,r}{2\sqrt{2}}\left(\frac{1}{m}+\frac{1}{\mu}\right),\quad S=2\cr
\eln

   These amplitudes lead to a new term to the amplitudes given in (6.17).
They are
\bln{ax5}
	\bar g^{(1)}_{11}\,r\quad\mbox{and}\quad\bar g^{(1)}_{12}\,r
\eln
where $\bar g$'s are new parameters in the problem.  Note that 
the decay-coupling amplitudes $F^J$ for $\ell=1$ are proportional to $r^2$, and hence
they violate the rule that we have adopted in this paper, namely that the $F^J$ for 
a given given $\ell=2$ are always proportional to $r^\ell$.

   We now consider another additional decay amplitude which have not been included in the
this paper.  It comes from allowing more than one factor of $p$ to be used in constructing
decay amplitudes.  Consider
\bln{ax6}
   A^J_{\ell\,S}(S\delta)=p_\mu\,p_\alpha\;\psi^{\mu\alpha}(S\,\delta)\;
	\,\chi^\beta(\ell\,0)\,\phi^*_\beta(J\,\delta)
\eln  
In the $J$RF, this leads to
\bln{ax7}
   A^J_{\ell\,S}(S+)&=0\cr
   A^J_{\ell\,S}(S0)&=(10\;10|S0)\,w^2\,(\gamma_s\beta_s)\,(-\gamma_\sigma\beta_\sigma)
	=-(10\;10|S0)\,\left(\frac{w^2}{4m\mu}\right)r^2\cr
	&=\frac{1}{\sqrt{3}}\,\left(\frac{w^2}{4m\mu}\right)r^2,\quad S=0\cr
	&=0,\quad S=1\cr
	&=-\sqrt{\frac{2}{3}}\,\left(\frac{w^2}{4m\mu}\right)r^2,\quad S=2\cr
\eln
These amplitudes lead to a new term to the amplitudes given in (6.17).
They are
\bln{ax8}
	\bar g^{(1)}_{10}\,\left(\frac{w^2}{m\mu}\right)r^3
	\quad\mbox{and}\quad
	\hat g^{(1)}_{12}\,\left(\frac{w^2}{m\mu}\right)r^3
\eln
where $\bar g$ and $\hat g$ are new parameters in the problem. 
Again these decay amplitudes have not been included in the Section 7 of this paper,
because the decay-coupling amplitudes $F^J$ for $\ell=1$ are proportional to $r^3$.

We consider the additional decay amplitudes considered in this section to be `anomalous.' 
They are highly relativistic, as they vanish faster than those treated in the Section 7
as $r\to0$.  Conversely, they may dominate the decay amplitudes as $w\to\infty$
and therefore must be included in any general treatment of the problem.

   There exist two additional anomalous amplitudes for the decay
\bln{ax9}
	2^-\to1^++1^-,\quad\ell=2
\eln
One of them is, evidently,
\bln{ax10}
    A^J_{\ell\,S}(S\delta)=\epsilon_{\mu\,\alpha\,\beta\,\gamma}\;\psi^{\mu\alpha}(S\,\delta)\,
	\chi^{\beta\rho}(\ell\,0)\,\tilde g_{\rho\tau}(w)\,{\phi}^{*\,\tau\gamma}(J\,\delta)
\eln
It takes on the form, in the $J$RF,
\bln{ax11}
   A^J_{\ell\,S}(S\delta)&=\sum_{m_a\,m_b}\,(1m_a\;1m_b|S\delta)\;
	\epsilon_{\mu\,\alpha\,\beta\,\gamma}\,	
	\omega^\mu(s\,m_a)\,\varepsilon^\alpha(\sigma\,m_b)\,
	\chi^{\beta\rho}(\ell\,0)\,\tilde g_{\rho\tau}(w)\,{\phi}^{*\,\tau\gamma}(J\,\delta)\cr
    &=\sum_{m_a\,m_b}\,(1m_a\;1m_b|S\delta)\cr
	&\hskip60pt\times\Bigg\{\omega^0(s\,m_a)
		\Big[\varepsilon(\sigma\,m_b)\,\chi(\ell\,0)\,\cdot\,\phi^*(J\,\delta)\Big]\cr
	&\hskip140pt-\Big[\omega(s\,m_a)\,\chi(\ell\,0)\,\cdot\,\phi^*(J\,\delta)\Big]
					\varepsilon^0(\sigma\,m_b)\Bigg\}\cr
\eln
so that
\bln{ax12}
   A^J_{\ell\,S}(S2)&=0\cr
   A^J_{\ell\,S}(S1)&=\frac{1}{3\sqrt{2}}\sum_{m_a\,m_b}\,(1m_a\;1m_b|S1)\cr
	&\hskip60pt\times\Bigg\{\omega^0(m_a)
		\Big[\varepsilon(m_b)\,
	\left(\chi(-)\,\phi^*(0)+2\chi(0)\,\phi^*(+)\right)\Big]\cr
	&\hskip100pt-\Big[\omega(m_a)\,
	\left(\chi(-)\,\phi^*(0)+2\chi(0)\,\phi^*(+)\right)\Big]
					\varepsilon^0(m_b)\Bigg\}\cr
&=\frac{1}{6}\,(-)^S\Bigg\{\omega^0(0)
		\Big[\varepsilon(+)\,
	\left(\chi(-)\,\phi^*(0)+2\chi(0)\,\phi^*(+)\right)\Big]\Bigg\}\cr
&\hskip100pt-\frac{1}{6}\Bigg\{\Big[\omega(+)\,
	\left(\chi(-)\,\phi^*(0)+2\chi(0)\,\phi^*(+)\right)\Big]
					\varepsilon^0(0)\Bigg\}\cr
&=\left(\frac{i}{3}\right)\Big[(-)^S\gamma_s\beta_s+\gamma_\sigma\beta_\sigma\Big]
	=\left(\frac{i\,r}{6}\right)\left[\frac{(-)^S}{m}+\frac{1}{\mu}\right]\cr
   A^J_{\ell\,S}(S0)&=0\cr
\eln
So the decay-coupling amplitudes $F^J$ for $\ell=2$ have a term proportional to $r^3$.
The second anomalous amplitude is
\bln{ax14}
   A^J_{\ell\,S}(S\delta)=p_\mu\,p_\alpha\;\psi^{\mu\alpha}(S\,\delta)\;
	\,\chi(\ell\,0)^{\nu\beta}\,\phi^*_{\nu\beta}(J\,\delta)
\eln  
which leads to, in the $J$RF,
\bln{ax15}
	A^J_{\ell\,S}(S2)&=A^J_{\ell\,S}(S1)=0\cr
	A^J_{\ell\,S}(S0)&=-\sqrt{\frac{2}{3}}\;(10\;10|S0)\,\left(\frac{w^2}{4m\mu}\right)r^2\cr
	&=\frac{\sqrt{2}}{3}\,\left(\frac{w^2}{4m\mu}\right)r^2,\quad S=0\cr
	&=0,\quad S=1\cr
	&=-\frac{2}{3}\,\left(\frac{w^2}{4m\mu}\right)r^2,\quad S=2\cr
\eln
Here the decay-coupling amplitudes $F^J$ for $\ell=2$ have a term proportional to $r^4$.
%
%
\section*{Acknowledgments}
\indnt
   S. U. Chung is indebted to the German Ministry of Science and Technology 
(BMBF) and the DFG for providing him with an appointment of Visiting
Professorship for the years 2003 through 2005 and again 
for the years 2007 through 2009.  This work started in 2003
during his visits to Bonn and Munich under the sponsorship.
He acknowledges with gratitude the financial support in 2007 as scientific consultant
to the Cluster of Excellence for Fundamental Physics in Munich, Germany.

He is also indebted to Larry Trueman/BNL and Ron Longacre/BNL 
for their numerous helpful comments during the final stages of this paper.

Finally, both of us acknowledge with pleasure the support and encouragement
from Professor Stephan Paul, Technical University Munich.
\brlist
\brf{SCh0} S. U. Chung, Phys. Rev.  D{\bf 57}, 431(1998).
\brf{SCh1} S. U. Chung, Phys. Rev.  D{\bf 48}, 1225(1993);
	Phys. Rev.  D{\bf 56}, 4419(1997).
\brf{SCh2} S. U. Chung, `Spin Formalisms,' CERN Yellow Report 71-8 (1971).
\brf{McK} A. McKerrell, Nuovo Cimento {\bf 34}, 1289 (1964).
\brf{Mcf} A. J. Macfarlane, J. Math. Phys. {\bf 4}, 490 (1963).
\brf{Rose} M.E. Rose, Elementary theory of angular momentum\\
 (John Wiley \& Sons, Inc., New York, 1957).
\brf{MG} M. Maggiore, `{\em A Modern Introduction to Quantum Field Theory},
	(Oxford University Press, Oxford, 2005), Chapter 2.
\brf{JW} M. Jacob and G. C. Wick, Ann. Phys. (N.Y.) {\bf 7}, 404 (1959).
\brf{RS} W. Rarita and J. Schwinger, Phys. Rev. \undl{60}, 61 (1941).
\brf{BrFr} R. E. Behrends and C. Fronsdal, Phys. Rev. \undl{106}, 345 (1957);\\
	C. Fronsdal, Nuovo Cimento Suppl. \undl{9}, 416 (1958).
\brf{Zm} C. Zemach, Phys. Rev. {\bf140}, B97 (1965); {\it ibid.} {\bf140}, B109 (1965)
\brf{Higgs}B. W. Lee, C. Quigg, and H. B. Thacker, Phys. Rev. D {\bf 16}, 1519 (1977); 
M.~S.~Chanowitz and M. K. Gaillard, Nucl. Phys. {\bf B261}, 379 (1985);
G.~J.~Gounaris, R.~K\"ogler and H.~Neufeld, Phys. Rev. D  {\bf 34}, 3257 (1986);
W.~Marciano and S.~Willenbrock, {\it ibid.} {\bf 37}, 2509 (1988).
\brf{BW} F. von Hippel and C. Quigg, Phys. Rev. \undl{5}, 624 (1972).
\erlist
\end{document}